# Daily Profile of COVID-19 Infections in Germany, throughout the Pandemic

Derek Marsh

Max-Planck Institute for Multidisciplinary Sciences[*], 37070 Göttingen, Germany.

[*]Formerly, Max-Planck-Institut für biophysikalische Chemie.

E-Mail: dmarsh@gwdg.de

## Abstract

Progress of the COVID-19 pandemic was quantified, in the first instance, using the daily number of positive cases recorded by the national public health authorities. Averaged over a 7-day window, the daily incidence of COVID-19 in Germany reveals clear sections of exponential growth or decay in propagation of infection. Comparing with incidence profiles according to onset-of-symptoms shows that reporting of cases involves variable delays. Observed changes in exponential rates $r$ come from growing public awareness, governmental restrictions and their later relaxation, annual holidays, seasonal variation, emergence of new viral variants, and from mass vaccination. Combining the measured rates $r$ with epidemiological parameters established for SARS-CoV-2 yields the dynamics of change in disease transmission. Combined with the distribution of serial intervals (or generation times), $r$ gives basic and instantaneous values of the reproduction number ($R_0$ and $R_t$, respectively) that govern development and ultimate outcome of the epidemic. Herd immunity requires vaccination of approx. 70% of the population, but this increases to ca. 80% for the more transmissible α-variant. Beyond this point, progressive vaccination reduces the susceptible population, and competes with the emergence of new variants. By the first Omicron wave, ca. 70% were doubly vaccinated, with the target then standing at ca. 80%. Combined with the distribution of times-to-death, incidence rates $r$ from onset of symptoms predict the daily profile of COVID-associated deaths and estimated case-fatality ratio. Cases are under-reported in the first wave and reflect age heterogeneity in fatalities at the second wave. In periods of low incidence, COVID mortality was ≲1% of detected infection.

## Introduction

The pandemic caused by SARS-CoV-2 brought with it profound medical, sociological, economic and political consequences for us all. It is likely to do so for quite some time to come, not least in retrospective analysis of the measures taken and in developing strategies to handle possible future pandemics. Following progression of the disease from day to day is therefore of major public concern. The primary source of information is the number of new positive cases registered daily by the national health authorities and reported publicly in various forms.

Straightforward appreciation of the daily recorded incidence in COVID-19 infection is hampered by a pronounced variation over the days of the week: the so-called "weekend effect". This problem is addressed variously in published data, e.g., by comparing with the same day one week previously, or by summing over the last seven days. Accepting retrospective analysis, a moving average centred on a 7-day window relates most directly to the date of original infection. Here, I define the average for day-$i$ as the sum over incidences from day-($i$-3) to day-($i$+3), divided by seven. It differs from the often-used 7-day incidence in that the origin is three days earlier (and expressed per day). Underlying trends that vary smoothly over a period of seven days are unchanged by this averaging process, as also are broad peaks. For instance, a Gaussian of width 6.5 days at half-height broadens only little, whereas that with 3.5 days starting-width approximately doubles in width; and a sharp singularity spreads uniformly over a 7-day period.

The daily profile produced by the moving average lets us identify regions of fixed exponential growth or decay and measure their rates. We expect exponentially varying changes when the number of infectious cases is small compared with the total susceptible population. This arises if transmission comes from random contacts between infectives and susceptibles; also recovery or death is often modelled as exponential (see e.g., Anderson and May, 1991; Hethcote, 2000). Points of change in exponential rate that arise from interventions introduced to control progress of the pandemic, and from their subsequent relaxation, also should be evident. The size of changes in rate then lets us estimate the effectiveness of control measures and predict final outcomes. Specifically, we can combine the exponential incidence rates - that directly characterise the dynamics of transmission - with epidemiological parameters such as distribution of generation times (or serial intervals) to determine the basic and instantaneous reproduction numbers, $R_0$ and $R_t$.

Here, I apply this approach to the daily case incidence of COVID-19 in Germany, reported to the Robert-Koch Institute (RKI, 2021a), which is the official organ for collating data on infectious diseases in the Federal Republic. (Appendix 1 gives an overview of various publicly available sources for data on COVID incidence in Germany at the beginning of the pandemic, and their inter-relation.) I begin with a summary of the theoretical background, before going on to compare data based on official reporting date with those based on date for onset of symptoms. The latter, together with distribution of serial intervals, translates to time evolution of the instantaneous reproduction number $R_t$. Similarly, combining with the distribution of onset time-to-death, yields predictions of the daily profile of COVID-associated deaths and, comparing with recorded deaths, estimates of case-fatality ratio.

Although begun during the pandemic, this account is retrospective. It therefore helps, to enumerate at the outset the stages through which the epidemic evolved. Table 1 lists the various phases into which the RKI proposed to divide the developing pandemic in Germany (Schilling et al., 2021; Tolksdorf et al., 2022). From initial outbreak, through summer troughs, come waves that are either seasonal, associated with the emergence of new dominant variants of the virus, or both. Classification by the RKI is based on a variety of factors: clinical and medical, in addition to epidemiological. As we see from Fig. 1, it strongly reflects reported daily cases.

**Mathematical Background**

*Compartmental Models and Exponential growth*

In the basic susceptible-infectious-removed (SIR) model (Kermack and McKendrick, 1927), the rate equations for the fractional population (or concentration) of susceptible and removed compartments, $[S] \equiv S/N$ and $[R] \equiv R/N$, are:

$$d[S]/dt = -\beta[S][I] \tag{1}$$

$$d[R]/dt = \gamma[I] \tag{2}$$

where $\beta$ is the rate of transmission by an infectious individual, and $\gamma$ is the recovery (or removal) rate of an infectious individual. The numbers of susceptible, infectious and removed individuals are $S$, $I$ and $R$; and $N = S + I + R$ is the total population. The right-hand side of Eq. 2 is the recovery rate of the infectious population $I$. Therefore, the time for recovery from infectiousness is distributed exponentially, with *mean* value $1/\gamma$ in this model (Hethcote, 2000). The second-order rate constant for transmission $\beta$ depends on the contact frequency of susceptibles with infectives multiplied by the probability of infection on contact, as in the law of mass action (de Jong et al., 1995; Hethcote, 2000). Later, we shall introduce distributions in generation time that characterize the transmissibility, $\beta$.

When the total population ($N$) remains fixed, the rate of increase in infectious population ($I$) is the difference between infection and recovery rates, Eqs. 1 and 2:

$$d[I]/dt = -d[S]/dt - d[R]/dt = (\beta[S] - \gamma)[I] \tag{3}$$

If the number of infections is low, relative to the population of susceptibles, $[S]$ remains approximately constant over a limited period. According to Eq. 3, the infectious population (and hence the daily fraction of new infections, $-d[S]/dt$ from Eq. 1) then grows exponentially with effective rate constant $r = \beta[S] - \gamma$. At the beginning of the epidemic, $[S]_0 \approx 1$ and the exponential rate constant is $r_0 = \beta - \gamma$. Correspondingly, the number of new infections decreases exponentially when $[S] < \gamma/\beta$ ($\equiv 1/R_t$, where $R_t$ is the instantaneous reproduction number).

*Basic Reproduction Number*, $R_0$

The basic reproduction number $R_0$ is the average number of new infections produced by a typical individual throughout its infectious lifetime, when the entire population is susceptible. Expressed *per capita*, the instantaneous rate of transmission is the number per unit time $n(\tau)$, where $n(\tau).d\tau$ is the number of infections produced by an individual in time interval $\tau$ to $\tau + d\tau$ after becoming infected. The reproduction number is the sum over all $\tau$:

$$R_0 = \int_0^\infty n(\tau).d\tau \tag{4}$$

Written in terms of the transmission rate $n(\tau)$, the probability density function $g(\tau)$ for the generation interval $\tau$ between primary and secondary infections is:

$$g(\tau) = n(\tau)/R_0 \tag{5}$$

where we use Eq. 4 for the normalizing denominator. The number of new infections at time $t$ is the sum of all infections caused by individuals infected at time $\tau$ ago (i.e., at times $t - \tau$). This results in the renewal equation:

$$C(t) = \int_0^\infty [S(t)]C(t-\tau)n(\tau).d\tau = R_0[S(t)] \int_0^\infty C(t-\tau)g(\tau).d\tau$$

(6)

where $C(t)$ is the count of new infections, and we use Eq. 5 for the right-hand side. Here, $R_0[S(t)]$ is the reproduction number at time $t$, where $R_0$ is the basic reproduction number at $t = 0$ when $[S(0)] = 1$.

We see from the renewal equation that the daily instantaneous reproduction number, $R_t \equiv R_0[S(t)]$, is the number of new infections $C_t$ at day $t$, divided by the total number of infectors causing these infections (Fraser, 2007):

$$R_t = \frac{C_t}{\sum_{i=1}^n C_{t-\tau_i} g_{\tau_i}}$$

(7)

where $\sum_{i=1}^n g_{\tau_i} = 1$, i.e., $n$ is the number of days over which the probability density for the generation time $g(\tau)$ is discretized. Generation times $\tau_i$ are always positive. However, if we use serial intervals as proxys, $\tau_i$ becomes negative whenever infectiousness precedes symptoms. The instantaneous $R_t$ defined in Eq. 7 gives the number of new infections produced by an individual infected at day $t$, if conditions remain those prevailing at $t$ (Fraser, 2007). It depends on the backwards directed distribution of generation times, because it derives from the number of secondary infections produced at day $t$, cf. the denominator in Eq. 7 (see also Gostic et al., 2020).

A simple example of instantaneous reproduction number is for a delta function distribution, $g(\tau) = \delta(\tau - T_G)$. Eq.7 then becomes:

$$R_t = C_t/C_{t-T_G}$$

(8)

where $T_G$ is the single unique generation time. The instantaneous $R_t$, for this case, is simply the ratio of daily new cases distanced $T_G$ days apart.

In regions where the rate of change in incidence varies exponentially, $C(t) = C_o \exp(rt)$, the renewal equation (Eq. 6) becomes (Wallinga and Lipsitch, 2007):

$$\frac{1}{R_0} = \int_0^\infty e^{-r\tau} g(\tau).d\tau$$

(9)

Then $1/R_0$ is the Laplace transform of the generation-time probability density $g(\tau)$, with respect to exponential rate constant for infection $r$. This corresponds to the Lotka-Euler equation in demography, where $r$ is the Malthusian parameter. Eq. 9 applies rigorously to generation times, and to probability densities that do not extend below $\tau = 0$. For distributions extending below $\tau = 0$, we should extend the lower integration limit to negative values.

As noted already, the distribution of infectious lifetimes decays exponentially in the SIR model. The normalized probability density is:

$$g(\tau) = \gamma \exp(-\gamma\tau) \tag{10}$$

when $\gamma$ is fixed. The mean lifetime is given by: $\langle\tau\rangle = \int \tau g(\tau) d\tau = 1/\gamma$. For constant transmissibility $\beta$, Eq. 10 also becomes the distribution in generation times. Taking the Laplace transform, the reproduction number for an exponential distribution of generation times is:

$$R_0 = 1 + r\overline{T}_G \; (= \beta/\gamma) \tag{11}$$

where $\overline{T}_G \equiv 1/\gamma$ is the mean generation time, and for the SIR model: $r = \beta - \gamma$ (see Eq. 3). For this particular case, the reproduction number depends linearly on the rate constant $r$. However, the gamma distribution offers a more realistic probability density function for the generation time, or serial interval as proxy (Bi et al., 2020; Cereda et al., 2020):

$$g(\tau) = \frac{(m\gamma)^m}{\Gamma(m)} \tau^{m-1} e^{-m\gamma\tau} \tag{12}$$

where $\Gamma(m)$ is the gamma function. The mean is again $\overline{T}_G = 1/\gamma$, and the standard deviation is $SD = 1/(\gamma\sqrt{m})$. This distribution is equivalent to $m$ successive exponential stages each of duration $1/(\gamma m)$ (Lloyd, 2001). From the Laplace transform, the reproduction number is:

$$R_0 = (1 + r\overline{T}_G/m)^m \tag{13}$$

where we get $m$ from the standard deviation. Sometimes, a Gaussian distribution is appropriate because this allows negative values of the serial interval, which correspond to onset of infectiousness before that of symptoms (see Du et al., 2020; Ali et al., 2020; Ferretti et al., 2020). The reproduction number is (Marsh, 2025):

$$R_0 = \frac{1 - \Phi((\tau_m - \overline{T}_G)/\sigma)}{1 - \Phi((\tau_m - \overline{T}_G)/\sigma + r\sigma)} \exp(r\overline{T}_G - \tfrac{1}{2}r^2\sigma^2) \tag{14}$$

where $\sigma \equiv SD$ is the standard deviation; $\tau_m \leq 0$ is the lower limit for the integrals in Eqs, 4,6,9 (matching that for the sum in Eq. 7), which we include to allow negative SIs; and $\Phi(x') = \int_{-\infty}^{x'} \exp\left(-\tfrac{1}{2}x^2\right). dx / \sqrt{2\pi}$ is the cumulative normal distribution function up to $x = x'$. Eq. 14 holds for generation times, which are always positive and $\tau_m = 0$. When using serial intervals as proxy, the lower limit of the integral in Eqs. 4 and 9 extends, in principle, to $-\infty$, which yields $\Phi(-\infty) = 0$ in Eq. 14 and hence:

$$R_0 = \exp(r\overline{T}_G - \tfrac{1}{2}r^2\sigma^2) \tag{15}$$

This is the result usually quoted for a Gaussian distribution (e.g., Wallinga and Lipsitch, 2007; Du et al., 2020; Ganyani et al., 2020), although it should only be used with very narrow distributions, i.e., for $\sigma \to 0$.

For a single unique generation interval $T_G$, the distribution is a delta function $g(\tau) = \delta(\tau - T_G)$, and similarly for $\beta(\tau)$ because infection occurs only at the single unique event.

Essentially, this is a fixed latent period followed by immediate infection. From Eq. 9, the reproduction number is then simply:

$$R_0 = \exp(rT_G) \tag{16}$$

This is the ratio of daily new cases distanced $T_G$ days apart in the exponential regime, as given more generally by Eq. 8 above. The delta probability density yields an upper bound for $R_0$, relative to generation-time distributions with non-vanishing width and the same mean value (Wallinga and Lipsitch, 2007). When using Eq. 8 for non-singular distributions, however, we might approximate $T_G$ better by taking the peak in generation-time distribution, instead of the mean.

*Outcomes in the SIR Model*

We deduce final outcomes of an epidemic by taking the limit $t \to \infty$ when solving Eqs. 1, 3 from the SIR model. From these two equations we get $d[I]/d[S]$, which integrates to:

$$[I] = [I]_o + [S]_o - [S] + (\gamma/\beta)\ln([S]/[S]_o) \tag{17}$$

where subscripts refer to $t = 0$, e.g., $[S]_o \equiv [S(0)]$. Taking initial conditions $[I]_o \cong 0$, i.e., $[S]_o \cong 1$, and final condition $[I]_\infty = 0$ as $t \to \infty$ , we get the total fraction of infections by the end of the outbreak, $\rho \equiv 1 - [S]_\infty$, from:

$$1 - \rho = \exp(-R_0\rho) \tag{18}$$

where $R_0 = \beta/\gamma$ in the SIR model of Eqs. 1–3. This is the final-size equation. At this point, we reach herd immunity. Earlier in the outbreak, the infectious population reaches a maximum $[I]_{max}$, before decaying to zero as $t \to \infty$ . Without preventative interventions, the maximum occurs when $d[I]/dt = 0$, for which $[S] = \gamma/\beta = 1/R_0$ (see Eq. 3). Using Eq. 17, we then get:

$$[I]_{max} = 1 - (1 + \ln(R_0))/R_0 \tag{19}$$

with the initial conditions given previously. Note that the final size deduced from Eq. 18 applies more generally than the SIR model; amongst other possibilities this includes both non-zero latency and general distributions of infective period (Ma and Earn, 2006).

*Outcomes in Probabilistic Models*

Alternatively, we can replace the deterministic SIR model by using probabilistic descriptions of the initial outbreak as a branching process. For a gamma distribution, the probability $\Pi$ of developing a major outbreak (i.e., exponential growth) is the solution to the balance equation (Anderson and Watson, 1980; Britton and Lindenstrand, 2009):

$$1 - \Pi = (1 + \Pi R_0/m)^{-m} \tag{20}$$

where $m$ is as defined in Eq. 12, and $R_0$ is given by Eq. 13. The gamma distribution approximates the serial intervals that one finds with COVID infections (Li et al., 2020; Bi et al., 2020; Cereda et al., 2020). Note that as $m \to \infty$, Eq. 20 tends in the limit to $1 - \Pi = \exp(-R_0\Pi)$, corresponding to fixed infectivity (up to $\tau = 1/\gamma$, see Eq. 12). On the other hand, $m = 1$ (i.e., an exponential distribution - see Eq. 12) yields $\Pi = 1 - 1/R_0$. The mean final size of the outbreak is $\rho$ conditional on the occurrence of a major outbreak, which for large populations ($N \to \infty$) is close to the solution of Eq. 18, as expected for a central limit (Britton and Lindenstrand, 2009; Anderson and Watson, 1980).

**Daily Rate of incidence**

Figure 2 shows the daily incidence of new confirmed cases in Germany, starting from 1 Mar 2020 as day-1. For the moment, we restrict ourselves to the period of the initial outbreak and its reduction by non-pharmaceutical interventions (1 Mar to 9 Aug). This comprises the first wave (I) and subsequent summer trough (a). The upper panel of Fig. 2 uses the official date of reporting to the local health authorities, whereas the time axis in the lower panel is that for onset of symptoms or first diagnosis. Data for the latter is incomplete, amounting to 71.4% (SD 1.3%) of total reported cases when averaged over the slowly varying range from day-64 to day-162. Solid triangles are individual daily case numbers, and open circles are averages over a 7-day window centred on each day plotted. We see a clear weekly periodicity of reporting in the upper panel that the moving average filters out. (Fourier transform of the raw data, up to day-1188, reveals prominent peaks at frequencies: 0, 1/7, 2/7 and 3/7 day$^{-1}$, with magnitude ratios 1.00:0.42:0.10 relative to the first harmonic.) As we might expect, such a periodicity is less evident in the data based on symptoms onset given in the lower panel. Individual maxima appear on some Mondays in the onset data (particularly evident here at times of high incidence); presumably these correspond to peaks in diagnosis and appear with greater regularity at Mondays later in the pandemic. Irrespective of this, we must perform 7-day averaging on the symptoms-onset data, because the averaging window smooths out all fluctuations that occur rapidly within the 7-day timescale. The 7-day moving average is directly proportional to the RKI weekly incidence per 100,000 population, where the latter is shifted by three days to later times. The initial target upper ceiling of this regulatory metric (50 per 100,000/week) is shown as a horizontal dashed line in Fig. 2. It was soon exceeded, and ultimately reached a maximum of almost 2,000.

Both profiles of incidence in Fig. 2 are characterized by an initial rapid rise, and a maximum followed by a slower decrease to a low basal level. Towards the end of the period shown, the low rate of incidence begins to increase once again, as we shall see later. Each profile also contains a local singularity in basal level at around day-107 (Jun 15) that corresponds to severe local outbreaks of infection in the meat-processing industry and centres of high-density housing. The reporting data lags behind that from onset of symptoms by ca. 5 days. Maximum incidence in reporting occurs later than that in onset of symptoms. However, this does not represent a constant delay: the shapes of the two maxima are very different. Evidently, the distribution in reporting delay time distorts the profile of incidences for reporting. Onset of symptoms lies closer to the date of original infection: it depends on the distribution of incubation times, which has a mean value of 5-6 days (Bi et al., 2020; Lauer et al., 2020; Lau et al., 2021). Therefore, the onset-of-symptoms profile better reflects the progress of infection. Note that imputing the missing onset-of-symptoms data at this stage does not greatly change the overall shape of the profile based on onsets (RKI, 2020). We show this as grey symbols in the lower panel of Fig. 2.

Figure 3 shows the amplitude of the weekly modulation obtained by dividing the original data by the 7-day average. We approximate this by an absolute sine function:

$$(modln)_t \equiv (C_t - <C_t>)/<C_t> = A \times (|\sin(\pi(t - t_o)/wk)| - f_o) \qquad (21)$$

introduced originally by Dehning et al. (2020a) in a multi-parameter Bayesian analysis. Over the range up to day-297, the parameters from non-linear least-squares fitting of Eq. 21 are:

$A = 0.84 \pm 0.03$, $t_o = 0.95 \pm 0.04$ day, $wk = 7.00 \pm 0.002$ day and $f_o = 0.632 \pm 0.008$. The zero-level for weekly modulation, $f_o$, is close to the mean value of the absolute sine function, viz., $2/\pi = 0.637$ times the maximum amplitude. The normalised modulation amplitude reduces somewhat to $A = 0.74 \pm 0.05$, if we restrict fitting to the first 7 cycles up to day-52. This is comparable to a previous global Bayesian fit with the same data, viz. $A = 0.6$, CI [0.5,0.7] (Dehning et al., 2020b), that additionally contains effects of modelled interventions within this range.

After day-297, the modulation amplitude begins to increase, and the midweek maximum shifts from Thursday or Wednesday to Wednesday or Tuesday. Beyond day-733, the amplitude of the modulation increases abruptly with a midpoint around day-780 and transition width of ca. 20 days. This coincides with a large drop in the rate of testing at the end of the 2021-2022 winter wave of incidence (wave-V), as we see later in Fig. 7. This is also associated with a shift in maxima of the modulation away from Wednesdays to Tuesdays, and finally Mondays towards the end of 2022. See Appendix 2 for more details.

Public holidays are marked with asterisks in Fig. 2. We can expect dips in reporting here, followed by filling up with the delayed cases, analogous to the weekend periodicity of Fig. 3. This appears evident in the 7-day averages for reporting (upper panel of Fig. 1), where short plateau regions, following the Easter period and 1-May holiday, are preceded by sharper decreases (see also Fig. 4 immediately following). On the other hand, such irregularities are largely absent in the 7-day averages for onset-of-symptoms in the lower panel of Fig. 2. This suggests that the short public holidays had relatively minor effect on true incidence rates under preventative measures current at the time.

Figure 4 shows the 7-day average daily case numbers on a log scale, and extends beyond the time span of Fig. 2, up to the maximum of wave-IV. This plot reveals linear sections of the profile that represent regions of exponential growth or decay in transmission of infection with rate constants $r_0$ to $r_{14}$ given in Table 2. Rates for symptoms onset are similar to those for reporting, but ranges for reporting lag behind symptoms onset. For onset-of-symptoms (open diamonds), the initial and fastest region of growth (with exponential rate $r_0 = 0.26 \pm 0.01$ day$^{-1}$, i.e., doubling each 2.7 days) extends to day-10/11 (see also Fig. 2). Beyond this, the growth rate slows progressively, falling to zero at maximum incidence on day-16/17. Then follows an exponential decrease that changes to a faster rate ($r_2 = -0.06 \pm 0.001$ day$^{-1}$) at around day-32/33. These change points correspond to increased public awareness of the pandemic and accompanying governmental interventions, as discussed by Dehning et al. (2020a), Flaxman et al. (2020) and Brauner et al. (2021).

Mass gatherings were banned in Germany from 9-10 Mar; schools and day-care centres closed from 13 Mar; public spaces, including non-essential shops, bars and restaurants, and entertainment venues, were closed from 16 Mar; and general lockdown with advice to stay at home came on 23 Mar (ECDC, 2021). Allowing for the incubation time (with mean 5-6 days; Lau et al. 2021; Bi et al., 2020; Lauer et al. 2020) that precedes symptoms onset, the initial decrease in exponential growth rate of the onset-of-symptoms profile precedes the first restrictions and is the immediate response from growing public awareness. Maximum daily incidence and initial decay in the onset data correspond with the first two interventions, whereas the twofold acceleration in decay rate ($r_1$ to $r_2$) comes well after closure of public spaces but follows fairly closely on general lockdown. The distribution of delays in reporting, which we see when comparing upper and lower panels in Fig. 2, introduce additional complications that make corresponding comparisons with the profile from reporting dates rather difficult.

At day-59/60 for symptoms onset, the rate of decrease in incidence slows down ($r_2$ to $r_3$ in Table 2). This follows the first easing of restrictions: opening of small shops, bookstores and restricted areas in larger shops in mid-April, and precedes complete lifting of lockdown on 4-10 May, i.e., days-65 to -71 (ECDC, 2021). This point marks the end of joint federal restrictions: from then on different states began to act independently.

Incidence flattens off to a relatively low value at around day-94 for onset of symptoms (see Fig. 2). This corresponds to the "a" section of the 2020 summer trough specified in Table 1. Following the discontinuity at day-108, already mentioned, the incidence rate for symptoms onset changes from decreasing to increasing at day-126 (4 Jul; that for reporting date four days later). This is the "b" section of the summer trough; it corresponds to the start of school holidays, which are staggered between the different federal states and extend over the period from day-114 to day-196 (see horizontal bars in Fig. 4). Throughout this time there are several changes in exponential gradient, up to and including day-196. Three clearly separate sections appear in the symptoms-onset data; we list the rate, $r_4$, only for the first in Table 2. These sections approximate to three most populous groupings of states, which in order of holiday period are: Nordrhein-Westfalen; Niedersachsen with Sachsen-Anhalt, Sachsen and Thuringen; and Bayern plus Baden-Wuerttemberg. Distribution of delays produces less clear resolution in the reporting data, but with delay, corresponds roughly to that for symptoms onset.

The sharp rise of daily incidence beginning in October 2020 ($r_5$-range in Table 2), corresponds to a seasonal increase with onset of winter. This constitutes the second wave of the pandemic (W-II). We see the seasonal effect also from shifted profiles of initial daily incidence in the southern hemisphere, e.g., for Australia and South Africa, compared with the northern hemisphere (ECDC 2020). A pronounced localized dip in Fig. 4 characterizes delayed reporting over the Christmas/New Year period. This appears as a more muted plateau in the onset data.

At the beginning of 2021, the broad minimum in daily incidence that occurs over the range from day-347 to day-359 for onset of symptoms, and days 352-367 for reporting, corresponds to a change-over from dominance of the original SARS-CoV-2 variants to that of the newly appearing Alpha-variant (i.e., B.1.1.7). This is the start of wave-III. Similarly, the deep minimum in daily incidence at day-479 for onset of symptoms (day-486 for date of reporting) occurs where the Delta-variant (i.e., B.1.617.2) comes to dominate over the Alpha-variant. This is the start of wave-IV. We shall see this later in Fig. 6, where increasing incidence of the Alpha-variant intersects that of the previously existing variants between days-361 and -368, and incidence of the newly developing Delta-variant crosses that of the Alpha-variant between days-473 and -480.

## Basic reproduction number, R₀

The basic reproduction number is the average number of infections produced on introducing a single infectious individual into a homogeneous population of susceptibles. The epidemic grows when $R_0 > 1$, and declines when $R_0 < 1$. Thus $R_0$-values are important for assessing effectiveness of interventions and the extent of vaccination needed to halt the epidemic. For instance, the RKI uses the effective $R_t$-value for regression analysis to assess effectiveness of non-pharmaceutical interventions during the pandemic (an der Heiden et al., 2023).

We get $R_0$ from values of the initial exponential growth rate $r_0$, by using additional data for the distribution of generation times $T_G$ (see Eqs. 11, 13–16). Because $T_G$ is not directly accessible, we use serial intervals (SI) instead. These are the difference in times for onset of symptoms between primary and secondary infection. If incubation times are the same for primary and secondary cases, values of SI equal those of the generation period, simply shifted in time. A spread in incubation times increases that for the SI relative to $T_G$ (cf. Ganyani et al., 2020; Lehtinen et al. 2021). If we consider a cohort with symptoms onset at time $t$, forward-looking serial intervals are those corresponding best to the generation-interval distribution (Park et al., 2021). For $R_0$, we need serial intervals from early stages of the epidemic, because they decrease as protective measures are applied (Ali et al., 2020; Bi et al., 2020).

Within Europe, the mean SI for close contacts of positive cases is 6.6 days with standard deviation SD = 4.8 days, deduced using a gamma distribution from the early phase of the outbreak in Lombardy, Italy (Cereda et al., 2020). For early stages of the well-defined outbreak in Shenzhen, China, the mean SI is 6.3 days (SD = 4.2 days; gamma distribution) (Bi et al., 2020). Early, less extensive, data for Wuhan, China give mean SI = 7.5 days (SD = 3.4 days; gamma distribution) (Li et al., 2020). In the initial, pre-peak period of infection, the mean SI for mainland China excepting Hubei Province, is 7.8 days (SD = 5.2 days; Gaussian distribution), which decreases later as preventative interventions are undertaken (Ali et al., 2020). Over the entire range, from pre- to post-peak, the mean SI drops to 5.1 days (SD = 5.3 days), which is comparable to reports that assume a single distribution for all stages (see, e.g., Nishiura et al., 2020; Ferretti et al., 2020). Du et al. (2020) similarly report a mean SI of 3.96 days with SD = 5.3 days by fitting data for mainland China with an unchanging Gaussian distribution.

For gamma distributions of SI (Eq. 13), the basic reproduction number is $R_0 \cong 3.4 - 3.5$, when using SI-data from Lombardy (Cereda et al., 2020) and Shenzhen (Bi et al., 2020), and $R_0 \cong 5.2 - 5.3$ from the early Wuhan data where the SD is smaller and mean longer (Li et al., 2020). Correspondingly, we get $R_0 \cong 4.2 - 4.3$ for a Gaussian distribution (Eq. 14 and $\tau_m = -2$ days) with the pre-peak data for mainland China outside Hubei province (Ali et al., 2020), and $R_0 \cong 2.0$ for the entire dataset ($\tau_m = -5$ days). The latter compare with $R_0 \cong 1.7$ from Eq. 14 ($\tau_m = -5$ days) with SI data of Du et al. (2020) and unchanging Gaussian distribution. Note that the commonly used Eq 15 yields considerably lower values; e.g., $R_0 = 3.0$-3.1 and 1.4 for pre-peak and pre- to post-peak, respectively, of Ali et al. (2020). Here, we choose $\tau_m$ to match discrete $R_t$-calculations from the renewal equation (Eq. 7) with Gaussian probability density. The ranges quoted for $R_0$ embrace initial exponential rates $r_0$ from both reporting and onset data in Table 2. For comparison, Flaxman et al. (2020) estimate $R_0 = 3.8$ [CI: 2.4-5.6] from COVID mortality data when averaged over 11 European countries. For Germany, an der Heiden and Hamouda (2020) using Eq. 8 with $T_G = 4$ days estimate $R_0 = 3.3$ [PI: 3.2-3.4], and Dehning et al. (2020a) using the SIR model and effectively a recovery time of 8 days obtain $R_0 = 3.4$. Differing somewhat, Linka, Peirlinck and Kuhl (2020a) estimate $R_0 \cong 6.3 \pm 0.6$ for the initial stages of the outbreak in Germany. This is from a susceptible-exposed-infectious-removed (SEIR) model with latent and infectious periods of 2.5 and 6.5 days, respectively. Note that we expect higher values from the longer recovery time, and because the SEIR model gives intrinsically higher estimates than does the SIR model.

In contrast, Eq. 16 (for a delta-function) gives us the upper bound on reproductive number, relative to distributions with non-vanishing dispersion and the same mean $T_G$ (Wallinga and Lipsitch, 2007). Taking $T_G \cong 7$ days as representative (cf. mean SI-values above), we get: $R_0 \cong 6.2 - 6.4$ for the upper limit, from onset and reporting data, respectively. Of course, the delta-function distribution is a unique case. Realistically, it seems more appropriate to a sharp peak in infectivity, i.e., in $\beta(\tau)$, than to a single unvarying contact time $T_G$. For a gamma distribution (Eq. 12), the peak in infectivity occurs at infectious lifetime: $\tau_{max} = (1 - 1/m)/\gamma$. Using serial-interval distributions quoted above, we get $\tau_{max} = 3.5, 3.1$ and $6.0$ days (Bi et al., 2020; Cereda et al., 2020; Li et al., 2020), suggesting that $T_G = 4$ days is a suitable value to use in Eq. 8. This value is adopted by an der Heiden and Hamouda (2020) for their $R_0$-calculations, because the latent period is somewhat shorter than the incubation time, whose mean is 5-6 days (Bi et al., 2020; Lauer et al., 2020).

*Time-dependent Reproduction number*

Time evolution of the reproductive number for subsequent exponential phases $R_t$ proceeds as for $R_0$, using rate constants $r_t$ from Table 2. However, the instantaneous reproduction number $R_t$ (Eq. 7) records more continuously, by using individual daily case numbers $C_t$. Fig. 5 plots instantaneous $R_t$ as open triangles (SI from Bi et al., 2020), with corresponding exponential phases represented by horizontal bars (solid Eq. 13, same SI; dotted Eq. 16, $T_G = 4$ days). 7-day average case data for reporting (top panel) and onset (bottom panel) come from Figs. 1, 2, 4. For Eq. 7, we divide the gamma distribution $g(\tau)$ into $n = 18$ discrete one-day steps starting from $\tau = 0$.

The time-dependent reproduction number decreases rapidly from the initial value $R_0 \cong 3 - 4$, which has considerable spread because case numbers start low. It crosses $R_t = 1$ at the value of $t$ for maximum incidence of wave-1 in Figs. 2, 4, and falls to values of $R_t \cong 0.67 - 0.79$ for a prolonged period, on entering the summer trough. The localized discontinuity at $t = 102 - 104$ days, discussed already, appears as a sharp spike in $R_t$. This anomaly is followed again by a region with $R_t < 1$, until increasing incidence during school holidays reaches plateau values of $R_t \cong 1.25$, in section b) of the summer trough. The next peak at $R_t \cong 1.55$ arises from a sharp autumn increase in incidence that heralds seasonal wave-II of the epidemic. Then, we see an abrupt anomaly at the Christmas to New Year period, over days 301-313. Similar artefacts accompanying official holidays appear at days 400-408 (4-12Apr), 663-677 (23Dec-6Jan) and 777-784 (16-23Apr). These are more pronounced in reporting data than in symptoms onset (bottom panel). Beyond wave-II, further maxima in $R_t$ are associated mostly with progressive dominance of different CoV-2 variants: Alpha, Delta, and Omicron (see next section). Emergence of Delta (wave-IV) coincides with increased incidence accompanying the 2021 school summer holidays (cf. Fig. 4) giving the sharp rise in $R_t$ around day-491. After follows a seasonal autumn/winter peak from day-590 onwards, where Delta remains the dominant variant. Finally, subsequent maxima in $R_t$ correspond to various dominant Omicron variants, BA.1, BA.2, BA.5; and later BF.7, BQ.1, XBB1.5, beyond the range of Fig. 5.

Solid bars agree well with maxima and minima in daily trends because they use the same SI-data. Dotted bars correspond to the fixed $T_G = 4$ days used by the RKI (an der Heiden and Hamouda, 2020). Qualitatively, the latter approach follows the trends deduced from the more realistic SI-distribution (open triangles), but the maxima are lower and the minima higher. Quite generally, we expect smaller differences between the various models when absolute incidence rates $r_t$ decrease bringing $R_t$ closer to one (Wallinga and Lipsitch, 2007).

Solid circles in Fig. 5 show time dependence of the cohort reproduction number $R_t^C$. Unlike the instantaneous $R_t$ (which derives from the average number of secondary infections produced at time $t$), the cohort – or case – $R_t^C$ is the average number of individuals that a member of the cohort infected at time $t$ is likely to infect in the future. We get $R_t^C$ by summing values of the instantaneous $R_t$ weighted by the generation-time probability density, from time $t$ onwards (Fraser, 2007; and cf. Wallinga and Teunis, 2004):

$$R_t^C = \sum_{i=1}^{n} R_{t+\tau_i} g_{\tau_i} \tag{22}$$

where $n$ again is the number of time-points over which $g(\tau)$ is discretized (cf. Eq. 7). Weighting with $g_{\tau_i}$ smooths and broadens the instantaneous $R_t$, as we see in both panels of Fig. 5. Particularly, the sharp artefacts from delayed reporting at public holidays are largely suppressed in $R_t^C$. Peaks and troughs associated with the epidemic waves are reduced somewhat in the cohort $R_t^C$. But the clearest effect in both rising and falling phases is a shift to shorter times by about 6 days, close to the mean SI. We expect this for the forward-looking cohort $R_t^C$ because it anticipates coming increases and decreases in $R_t$ (cf., Cori et al., 2013). Note, however, that the instantaneous $R_t$ tracks the epidemic in real time, whereas we can only evaluate the cohort $R_t^C$ retrospectively (see, e.g., Gostic et al., 2020). Nevertheless, the comparison in Fig. 5 indicates likely biases in real-time estimates from the instantaneous $R_t$.

We turn now to the relation between basic reproductive number and rate constant $\beta$ for transmission of infection (see Eqs. 1, 3 and 11). The rate constant, $\gamma$, for removal of infectious individuals is the reciprocal of the mean SI. This gives $\gamma = 0.15 - 0.16$ day$^{-1}$, i.e., a half-time for recovery of 4.6-4.4 days (Cereda et al., 2020; Bi et al., 2020), from fitting with a gamma-distribution. Alternatively, we get $\gamma = 0.13$ day$^{-1}$ and half-time: 5.4 days from the Gaussian-distribution treatment of the pre-peak by Ali et al. (2020). Using values of $R_0$ deduced with the gamma distribution (viz., $R_0 = 3.4 - 3.5$), the transmission rate constant then becomes: $\beta = R_0 \gamma = 0.53 - 0.54$ day$^{-1}$ ($1/\beta = 1.9 - 1.8$ days) from onset and reporting data, respectively. Correspondingly, we get $\beta = 0.55$ day$^{-1}$ ($1/\beta = 1.8$ days), from the Gaussian-distribution approach for the pre-peak. For comparison, Dehning et al. (2020a) derive $\beta = 0.41$ [0.32,0.51] day$^{-1}$ from Bayesian inference with the SIR model and an informative prior of $\gamma = 0.12$ day$^{-1}$.

**New CoV-2 Variants**

As the pandemic progressed, SARS-CoV-2 variants of concern evolved that are more infective, or able better to evade the immune system. The top panel of Fig. 6 shows progressive growth and decay of different variants as new ones come to replace those previously dominant, starting from January 2021. Here, 7-day averages for numbers of daily cases come from weekly values provided by random sampling of positive cases with fully determined genomic sequence (RKI 2021c). Percentage populations are relative to total numbers sequenced. The bottom panel of Fig. 6 gives the ratio of positive cases for the new variant, relative to the previous one. In both panels, the *y*-axis is logarithmic. First comes the Alpha-variant (B.1.1.7), relative to the original strain, followed by Delta (B.1.617.2), then Omicron (B.1.1.529, alternatively denoted BA.1), and finally Omicron variants BA.2 and BA.5. Change over between variants occurs at days 355−376, 467−481, 663−677, 719−733 and 824−838, respectively. Growth and decay approaches exponential in these crossover regions, as we see more clearly from the ratios of the two coexisting variants in the lower panel.

Table 3 gives exponential rate constants of the dominant variants, $r_{new}$ and $r_{old}$, obtained from linear regression in the crossover regions of Fig. 6 (top). We can compare best here, where the two populations are of comparable sizes. Because total numbers sequenced varies, it is important that regression is over identical ranges for the two variants. For example over days 355−376, $r_{Alpha} = 0.031 \pm 0.003$ and $r_{orig} = −0.030 \pm 0.003$ day$^{-1}$, for Alpha and original variants, respectively. Also included in Table 3 are differences in rate constants, $\Delta r \ (\equiv r_{new} − r_{old})$, that we get from linear regression over slightly longer ranges in the bottom panel of Fig. 6. The differences agree with the individual values from the top panel, and precision is comparable.

When incidences of both old and new variants are exponential over the crossover region, we can use individual rate constants in Table 3 to estimate reproduction numbers from Eq. 9, assuming that generation-time distributions do not change appreciably. With the serial-interval distribution from Bi et al. (2020), instantaneous reproduction numbers on crossover to the Alpha variant are: $R_{t,\alpha} = 1.20 \pm 0.02$ and $R_{t,orig} = 0.82 \pm 0.02$. This translates to an increased transmissibility (cf. Eq. 11): $R_{t,\alpha}/R_{t,orig} \ (\equiv \beta_\alpha/\beta_{orig}) = 147 \pm 5\%$ for Alpha, relative to the original SARS-CoV-2 variants. If instead we calculate reproductive numbers from Eq. 8 or 16, with $T_G = 4$ days (as done by the RKI), Alpha transmissibility increases by $127 \pm 3\%$ relative to the original variants. Correspondingly, transmissibility predicted for Delta relative to Alpha is: $R_{t,\delta}/R_{t,\alpha} \ (\equiv \beta_\delta/\beta_\alpha) = 242 \pm 29\%$ with serial interval from Bi et al. (2020), and $177 \pm 15\%$ with the RKI approach. The difference between the two estimates reflects the growing disparity in $R_t$-values as these deviate from unity. Similarly, transmissibility of Omicron (BA.1) relative to Delta becomes: $R_{t,o}/R_{t,\delta} \ (\equiv \beta_o/\beta_\delta) = 249 \pm 39\%$ with serial interval from Bi et al. (2020), and $179 \pm 21\%$ with the RKI approach. Finally, for the Omicron variants, we get ratios: $R_{t,BA.2}/R_{t,BA.1} = 165 \pm 8\%$, $R_{t,BA.5}/R_{t,BA.2} = 204 \pm 13\%$ with Bi et al. (2020), and $137 \pm 4\%$, $156 \pm 7\%$ with the RKI approach.

For comparison, Davies et al. (2021) estimate $R_{Alpha}/R_{other} = 143−190\%$ (CI: 130-230%) for Alpha in England, by using a variety of statistical means. The odds for transmission of the Delta-variant relative to the Alpha-variant is 1.70:1 (95% CI 1.48−1.95) from statistical estimates for household clusters in England (Allen et al., 2022). Additionally, overall transmissibility of Delta is estimated as 2.1 times that of Alpha (95% CI 1.3−3.3) for UK households (Hart et al., 2022). These values are similar to those deduced here for Germany.

By multiplying relative transmissibilities, we estimate that the Delta-variant is around 3.0−4.1 times more infectious than the original SARS-CoV-2 variants (2.0−2.5× using the RKI estimates), and Omicron is 6.4−11.9 times more infectious than the original variants (3.2–5.0× using the RKI estimates). However, such extrapolations are primarily of academic interest.

We should remember that reproduction numbers of new variants that we estimate above are those prevailing at the time of changeover between dominant variants, and not those applying in the absence of vaccination, immunity gained on recovery, and preventative measures still in place. This holds also for the relative infectivities, although we might expect some cancellation on taking ratios (e.g., in $\beta_\delta/\beta_\alpha$). Also, reproduction numbers will change if generation-time distributions differ between variants. A statistical model for household transmission data in the UK reveals a decrease in mean generation time from 5.5 days for the Alpha variant to 4.7 days for the Delta variant (Hart et al., 2022). Using our method above would then overestimate the $R_{t,\delta}/R_{t,\alpha}$ ratio.

**Testing rates**

Clearly, if we increase the number of individuals tested, the number of positive cases also will increase, except when no new infections at all occur. Therefore, the growth and decay of positive daily cases shown in Figs. 2, 4 and 6 inevitably include any changes in rate of testing.

If testing rates change appreciably, this must be reflected by changes in $N_t$, the effective total population surveyed. The number of daily positive cases $C_t$ equals the daily decrease in susceptible population: $-\Delta S_t$, at day $t$. To allow for changes in testing numbers, we normalize $C_t$ to $N_t$ (cf. Eqs. 1–3). Taking the effective population as directly proportional to number of individuals tested $N_{test}$, we arrive at $C_t/N_{test}$. The corresponding day-by-day rate of change is:

$$\frac{d(C_t/N_{test})}{dt} = \frac{1}{N_{test}}\frac{dC_t}{dt} - \frac{C_t}{N_{test}^2}\frac{dN_{test}}{dt} = r_I\left(\frac{C_t}{N_{test}}\right)$$

(23)

where the equality on the right defines $r_I$. This is the epidemiologically significant rate constant. Hence, in regions of exponential change (i.e., fixed $r$), the rate constant for the fraction of daily infections $C_t/N_{test}$ becomes:

$$r_I = r - r_N$$

(24)

where $r$ is the usual rate constant for daily case numbers, and $r_N \equiv N_{test}^{-1}\, dN_{test}/dt$ is the effective rate constant for $N$. Increasing the testing rate $dN_{test}/dt$, whilst the intrinsic infection rate $r_I$ stays constant, therefore increases the number of daily reported new cases, as expected.

Independent of SIR or other compartmental models, the fraction of tests that register positive, $C_t/N_{test}$, is a useful empirical indicator of how much the rate of daily incidence is affected by changes in testing rate. If both daily incidence and number of tests per day change exponentially, then so must $C_t/N_{test}$, with a rate constant that is given by Eq. 24.

Figure 7 shows data on testing reported to the Robert-Koch Institute (RKI, 2021d; 2023c). Reporting was voluntary and it took some time to establish a stable number of contributing laboratories. The total number of tests reported $N_{test}$ is given by squares; the number of these found positive is given by circles, and the fraction of positive tests by triangles. Data extends up to 31 Jan 2023; beyond this point the number of reporting laboratories abruptly more than halved. The number of positive cases in Fig. 7 is less than given in Figs. 2 and 4, because data on testing is not available for all cases reported; with an additional complication that any particular individual may have more than one test. Also, the actual date of testing differs from that for statutory registration of positive cases with the local health authorities. Regions where exponential changes in daily positive cases correspond with approximately exponential change in daily number of tests are shaded grey in the log-linear plot. In these regions, the fraction of positive tests, $C_t/N_{test}$, also changes exponentially. We compare rate constants $r_I$, $r_N$ and $r$ for these three quantities in Table 4. Values of $r_I$ agree with the prediction for exponential rates given by Eq. 24. Up to day-530, the average rate constant associated with testing $r_N$ is 18% of $r$, although it reaches 33% at one point. Over this range, we expect true changes in incidence to account for most of the rate changes found in Figs. 2, 4 and 6. Beyond this, changes in rate of testing are considerably higher: on average 60% of the $r$-value for positive cases, over the range day-600 to -950. At this stage in the pandemic, data on both rates of incidence and absolute numbers of cases are distorted strongly by insufficient testing.

The increase in testing rate from day-600 coincides with onset of the 2022-autumn/winter wave of the pandemic, where the Delta-variant dominated (cf. Tolksdorf et al., 2022). This continues with the immediately following Omicron-wave, which starts from day-670 and attains the highest positive case numbers accompanied by maximum number of daily tests. Decay of the Omicron-wave from day-747 follows appearance of the BA.2 Omicron variant and a rapid reduction in number of daily tests. The latter arises from relaxing the requirements for a negative test to visit hospitals or care homes (from 16 Feb 22, day-718), to use public transport or attend indoor events (from 19 Mar 22, day-749), and dropping all restrictions including testing in schools (from 1 Apr 22, day762) (see Lionello et al., 2022).

Note that we can get a rough indication of $N$, the total effective population involved, by comparing with the maximum in infectious population, $I_{max} \equiv [I]_{max}N$, that Eq. 19 predicts for the SIR model. The infectious population on day-$n$ is the sum over all cases, $C_i$, not yet recovered/removed: $I_n = \sum_{i=n-T_I+1}^{n} C_i$. Here $T_I (= 1/\gamma)$ is the infectious lifetime, i.e., recovery time, which must be at least as long as the serial interval; we assume $T_I = 8$ days (cf. SI-values quoted previously). Using augmented onset-of-symptoms case data (RKI, 2021a), we then get $I_{max} = 39060$ at day-22 for the initial outbreak, and $R_0 = 3.18$ from the initial exponential rate ($r_0 = 0.273\pm0.009$ day$^{-1}$) with Eq. 11, leading to $[I]_{max} = 0.322$ from Eq. 19 and thus giving $N = 121200$ for the effective population sampled. This effective $N$ approaches 0.18% of the true population, which is the proportion estimated earlier for Germany by Linka et al. (2020b). Coincidentally, it is comparable to the number of weekly tests (ca. 129-374×10$^3$) reported at this stage of the pandemic (see Fig. 7). Of course, this value of $I_{max}$ (and hence $N$) is an underestimate because preventative measures were already introduced by this time. One way to look at this effective number is partly as a reduction in population of susceptibles by the non-pharmaceutical interventions. Note also that taking a longer $T_I$ period increases the estimated $N$.

**Age-stratified daily incidence**

Data presented so far refers to the whole population. Considering different social interactions and vulnerabilities, we do not expect incidence rates and overall infection necessarily to be the same for all age groups. Such issues are crucial to deciding upon preventative measures.

Figure 8 shows daily incidence based on reporting date for six different age groups. The top panel gives straight daily case numbers and includes data already presented for the entire population (squares). In the bottom panel, case counts are normalized to the fraction $N_j/N_{tot}$ of each age group $j$ in the total population. Although each follows the overall progression of the pandemic, incidence profiles for the different age groups do not all have the same shape. For instance, 0 to 4 year-olds and over 60 year-olds show little response to the school summer holidays; and the over 80 year-olds show a stronger maximum around Christmas 2020, but a reduced response to the following Alpha-variant wave.

After normalizing for total populations (Fig. 8, bottom), the first-wave peak is highest for the over 80 year-olds and much lower for the 0-4 and 5-14 year-olds. We might anticipate higher incidence for the elderly because of weaker immune systems and concentration in care homes. Relative values depend on testing regimes in the different age groups, but during wave-I testing was restricted mainly to suspected cases for all ages. Exponential growth rates $r$, on the other hand, have the advantage of not depending on absolute case levels. Reproduction numbers deduced from $r_t$ (with the SI distribution of Bi et al., 2020) therefore let us compare age groups at various stages of the pandemic. In wave-I, reproduction numbers $R_0$ are highest for 35-59 year-olds and above, decreasing with decreasing age below this (see Fig. A.3, in Appendix 3). Age differentials mostly become less pronounced in the subsequent stages, where values of $R_t$ are closer to one. Even so, some alternation in age dependence appears between different stages. The instantaneous $R_t$ for the entire population is the weighted average of the individual $R_{t,j}$ for each age group, $j$: $R_t = \sum_j N_j R_{t,j}/N_{tot}$. To within the expected uncertainty ranges, these values are close to determinations without age stratification shown in Fig. 5. For example, the basic reproduction number becomes $R_0 = 3.32 \pm 0.09$, as opposed to $R_0 = 3.37 \pm 0.08$ deduced from the initial exponential rate $r_0$ for the whole population.

## COVID-associated deaths

Data on fatalities following diagnosis of COVID, unlike case figures in the previous sections, do not suffer from possible under reporting. However, in not all cases is COVID necessarily the primary cause of death. We connect the day of death with incidence of COVID cases based on dates for onset-of-symptoms by using a probability distribution analogous to that introduced for the SI or generation time.

Open circles in the upper panel of Fig. 9 are weekly COVID-associated deaths given as a 7-day daily average centred on Wednesdays in the week of death. The initial peak lags behind that for COVID cases in Fig. 2 by approximately 20 days. Qualitatively, the subsequent development follows that in cases (cf. Figs. 1, 4): dropping to a low level and then increasing later in the year with the second seasonal wave of infection. Third and fourth waves follow corresponding to arrival of the Alpha and Delta variants, respectively, as already mentioned.

Given the probability density distribution $f^{(os-d)}(\tau)$ for time $\tau$ from onset-of-symptoms to death, we can predict daily death rates from the dependence of daily cases $C_{t_{os}}$ on symptom-onset date $t_{os}$ (see Fig. 2, bottom). The number $d_t$ of deaths on day $t$ is the sum over cases with symptoms-onset on previous days $t_{os}$, weighted by the probability $f_{t-t_{os}}^{(os-d)}$ that death occurs $\tau = t - t_{os}$ days after onset of symptoms (Flaxman et al., 2020):

$$d_t = (cfr)_t \sum_{t_{os}=0}^{t-1} f_{t-t_{os}}^{(os-d)} C_{t_{os}}$$

(25)

where $(cfr)_t$ is the daily case fatality rate (i.e., ratio of deaths to positive cases). Verity et al. (2020) fit early data for $f^{(os-d)}(\tau)$, from Hubei province and elsewhere in China, with a gamma distribution (Eq. 12) yielding mean onset-to-death $\bar{\tau} =$17.8 days and standard deviation 8.0 days (from SD/mean = 0.45). The top panel of Fig. 9 compares predictions (solid circles) with the observed mean daily rates calculated from weekly averages, centred on Wednesdays (RKI, 2021e). Predictions depend on the daily values of $(cfr)_t$ and also on degree of completeness of the onset data. Black circles in the top panel are predictions using straight onset data, whereas grey circles are augmented by imputed missing cases (RKI, 2021a). For ease of reference, we normalize predictions to the first peak in daily number of deaths. The probability density distribution from Verity et al. (2020) describes the shape of the first peak of the epidemic reasonably well, without further fitting. (Using the time course of COVID-associated fatalities reported in ECDC (2020) needs a longer onset-to-death, shifted to $\approx$ 25 days.) Comparing with incidence data for onset of infection in Fig. 2 (bottom) and Fig. 4, we see from Fig. 9 that the distribution $f^{(os-d)}(\tau)$ of onset-to-death times smears out the breakpoints in time course of daily case numbers. This agrees with an analysis of COVID-associated excess deaths in Europe by Flaxman et al. (2020), who conclude that non-pharmaceutical interventions in the first phase of the epidemic were too closely spaced to be resolved individually in data on fatalities.

With the normalization adopted in Fig. 9, predictions again come closer to observed fatalities towards the peak in second wave of the epidemic. This implies comparably high values of $(cfr)_t$ at this stage in the epidemic to those at the beginning. In the intervening region, predictions are much higher than actual deaths; correspondingly, values of $(cfr)_t$ are considerably lower. There are at least two contributions to decreased $(cfr)_t$ in the intermediate region: (i) under-reporting of positive cases at the beginning of the outbreak; and (ii) more cases in younger members of the population, for whom fatalities are less likely, as the second wave develops. In addition, seasonal variation in patterns of behaviour may also contribute.

The lower panel of Fig. 9 gives the time course for $(cfr)_t$ that comes from the ratio of observed deaths to those predicted from Eq. 25. We concentrate on the grey circles where missing onset data is included by imputation (RKI, 2020a). At the first peak in numbers of deaths, $(cfr)_t$ has the relatively high value of ca. 6%, which reduces to ca. 4% at the second peak, for reasons already given. In the intervening trough of low, slow-varying numbers of daily deaths, the mean value between days 130 and 221 is $\overline{cfr} = 0.0070$ (SD=0.0024). If we can neglect residual under-reporting, this translates to an intrinsic infection-fatality ratio (deaths to infections): $ifr = 0.7\%$. For comparison, Dimpfl et al. (2021) determine $ifr = 0.86\%$ [CI: 0.69-0.98] by combining RKI data with seroprevalence from the Schlogl outbreak in Austria. They explain why a previous value from the Gangelt outbreak (Streek et al., 2020) may be too low for Germany. Verity et al. (2020) find a *cfr* of 1.38% [CI: 1.23-1.53] for China, when adjusted for censoring, demography and under-reporting. For testing-intensive South Korea, the delay-adjusted *cfr* is 1.97% [CI: 1.94-2.00] (Shim, 2021). Additionally, linking age-stratified COVID-associated deaths in 45 countries with immune detection from 22 national-level surveys of seroprevalence yields a weighted average infection-fatality rate: *ifr* = 0.96% (O'Driscoll et al., 2021).

Comparison of these basal values with the two major peaks for $(cfr)_t$ in Fig. 9 indicates substantial under-reporting in the first two COVID waves. The mean value between days 11 and 102 for the first peak is $\overline{cfr} = 0.041$ (SD=0.013), and that between days 263 and 382 for the second is $\overline{cfr} = 0.031$ (SD=0.007). Attributing the peaks solely to unreported cases would imply that approximately 3/4 and 2/3 of cases escaped reporting at the peaks of first and second waves, respectively. By contrast, the peak corresponding to the third wave at around days 578 to 662 becomes close to basal: $\overline{cfr} = 0.009$ (SD=0.002). For comparison, Fiedler et al. (2021) estimate that a fraction 0.74 of cases go undetected in the first peak, around day-47 in Germany, and fraction 0.89 correspondingly in Italy.

Later in the pandemic, following an intensive vaccination programme, the values of *cfr* for Germany become much lower (see Fig. 9). For the range days 704-935 in 2022, the mean value reduces to $\overline{cfr} = 0.0013$ (SD=0.0003), at least part of which can be attributed to success of all preventive measures introduced by this stage. A further contributor is the reduced pathogenicity of Omicron variants, evident also in the hospitals, and a major contributor to recovery from the pandemic (see, e.g., Nyberg et al., 2022).

**Age-stratified COVID-associated deaths**

Figure 10 gives numbers of COVID-associated weekly deaths for different age groups averaged over different time segments of the pandemic. We normalize numbers to the fractional population of each age group: 0-4, 5-14, 15-34, 35-59, 60-79 and 80-plus years. The time segments correspond to the first wave, second wave plus Alpha-wave, Delta-wave plus first Omicron-wave (BA.1), subsequent omicron waves (BA.2, BA.5, etc.), and two further peaks of diminishing amplitude.

In common with overall deaths (solid circles, in Fig. 10), the population-corrected numbers for COVID-associated fatality decrease from birth to about ten-years old, and then increase exponentially from around 20 to 24-years of age onward. The exponential rate constant is $r \cong 0.11$ yr$^{-1}$ for most stages, corresponding to a doubling every 6.3 yr increase in age (for overall deaths in 2021, $r \cong 0.087$ yr$^{-1}$ and 8-year doubling). The increased COVID rate is particularly a hazard for the elderly, whose mortality rates are intrinsically higher.

**Vaccination**

Figure 11 gives the progress in vaccinating the population, from the time at which reliable vaccines first became available in Germany. The bottom panel, starting from the beginning of 2021, shows the percentage of the population vaccinated once, twice, and subsequently boosted for third and fourth times. Black squares in the top panel of Fig. 11 give the percentage of the total population assumed to have attained basic immunity, by being vaccinated twice or by recovery from infection before a single vaccination. The top panel also includes data for separate age groups, who were assigned different priorities in the vaccination programme.

At the time of the Alpha wave (around day-366 – see Table 3), little of the population in Germany was vaccinated. Within Europe, this variant was identified first in England (Davies, N.G., Abbott, S., et al., 2021). By the appearance of the Delta wave (day-476), more than half the population was at least singly vaccinated. Beyond this, vaccination may have contributed to evolutionary pressure on the virus.

Vertical dashed lines in Fig. 11 indicate dates at which much of the population had achieved basic immunization. By day-565, which is at the beginning of the autumn/winter stage in 2021, this corresponds to 63% of the whole population and 84% of the over 60 year-olds. The latter age group is particularly susceptible, and correspondingly was among the first vaccinated. Later by day-736, which follows the first (BA.1) Omicron wave, the percentages for the total population and over-60s increase to 76% and 89%, respectively. As we shall see immediately in the next section, these values fulfill the condition for herd immunity, based on transmissibility and basic reproduction number of the original CoV-2 virus.

## Prognoses

In the absence of interventions, final outcomes depend on the basic reproduction number $R_0$ that we determine at the beginning of the epidemic. If we describe the emerging outbreak as a branching process, the balance equation (Eq. 20) gives the probability $\Pi$ that a major outbreak develops. Taking $m = 2.25$ for the exponent in Eq. 20 (Bi et al., 2020), and correspondingly $R_0 = 3.46 \pm 0.01$, the probability of a major outbreak becomes $\Pi_0 = 0.85$. We can estimate the final fraction of infections that ensues from such a major outbreak by using the final-size equation, which is valid not only for the SIR model but also for a gamma distribution of infectious lifetimes (Ma and Earn, 2006). With the $R_0$ given already, Eq. 18 thus yields a value of $\rho_0 = 0.96$. Evidently, the probability of a major COVID-19 outbreak is theoretically very high, if preventative steps were not taken promptly. Otherwise, we can expect that nearly the whole of the population ultimately would have become infected. Prognoses become better when $R_t$ is reduced closer to unity, but these predictions apply only so long as the measures causing a decreased $R_t$ remain in place.

The basic reproduction number $R_0$ also determines how much of the population we must vaccinate to halt the epidemic. If we remove fraction $p$ of susceptible individuals by immunization, the effective reproduction number is reduced to $R = R_0(1 - p)$. The condition $R < 1$ for steady decay of the disease then leads to the critical proportion of the population that we need to vaccinate:

$$p_c = 1 - 1/R_0 \tag{26}$$

Taking $R_0$ from above, we find $p_c = 0.71 \pm 0.01$, for the critical fraction. As the pandemic proceeds, new variants that are more effective in transmission come to dominate, and also vaccination is introduced. Then we need to update our estimates correspondingly. At the time of the Alpha wave, vaccination levels were very low. The accompanying low instantaneous $R_t = 1.24$ (Fig. 5) resulted from preventative measures that mostly were impermanent. Hence, we must take the full transmission advantage of Alpha into account when using Eq. 26. From Fig. 6 above, we found that $R_{t,\alpha}/R_{t,orig} = 1.5$ for Alpha, relative to the original SARS-CoV-2 variants. This gives a corrected $R_0 = 5.1$ that increases the critical vaccination fraction to $p_c = 0.80$ for the Alpha variant.

The vaccination programme developed quickly, preventive restrictions were lifted accordingly, and reductions in $R_t$ relative to $R_0$ then reflected permanent effects of increased population immunity. If the fraction of the population fully vaccinated (or otherwise immune) is $v_t$ by time $t$, the fraction remaining that we need to vaccinate then is: $(1 - v_t)(1 - 1/R_t)$. At the beginning of the delta-wave around day-531, for example, 55% of the population were doubly vaccinated and the reproduction number based on reporting was $R_t = 1.43$ (Fig. 5). The further fraction of (double) vaccinations then required is 13%, totaling 68% in all, which is comparable to the original estimate from Eq. 26, but lower than the revised estimate at the time of the Alpha wave. Approaching the Delta peak (day-600), the new estimate increases to a total of 72%, whereas 63% was doubly vaccinated. By the approach to the first Omicron peak (BA.1) at day-678, the doubly vaccinated population increased to 69% and $R_t = 1.44$, increasing the vaccination target to 78%. Vaccination rates by then are beginning to level off, and no longer compete so efficiently with newly appearing variants. For approach to the BA.5 variant peak at day-830, $R_t = 1.54$ and 73% are double vaccinated, necessitating 9% further vaccination, i.e., an 82% target, well past the original estimate. This illustrates the interplay between vaccination rate and incidence rate. But we must remember that by choosing the peaks in $R_t$ these are definitely upper estimates. In between, the dynamics of infection are more forgiving, as illustrated by Fig. 12. This figure gives the projected total vaccination target:

$$p_{c,t} = 1 - (1 - v_t)/R_t^C \qquad (27)$$

as fraction, $v_t$, of the population vaccinated increases, starting from the Delta wave. We use the retrospective cohort $R_t^C$ as the best estimate, with data according both to reporting date and to onset of symptoms. Peaks in the projected target remain above 80% until after the BF.7 variant wave, and after the XBB.1.5 wave they fall to the 71% level predicted for the original CoV-2. By this time, the doubly vaccinated population has levelled off at 73-74%.

If the vaccine efficacy $e$ is less than unity, the critical fraction $v_c$ of the population that we must vaccinate increases such that: $p_c = e v_c$ in eqs. 26, 27. Efficacies for COVID-vaccines vary from 95% for mRNA vaccines, with mostly lower values for viral-vector vaccines, to 50% for inactivated virus (Mahase 2021). Strictly, the simple relation holds for all-or-none efficacy where fraction $e$ of the vaccinated becomes completely immune and the remainder receive no protection (Smith et al., 1984). At the opposite extreme (the "leaky" vaccine), everyone vaccinated has chance 1−$e$ of infection on encountering an infector. For the first round of encounters, fraction 1−$e$ become infected, the same as for all-or-none efficacy. However, those evading infection on the first round still have probability 1−$e$ of infection on the second and following rounds, whereas by this stage the remaining all-or-none vaccinees are all immune. Practically, this distinction also has implications for how we determine vaccine efficacy (Halloran et al., 1992). In Germany, vaccination was mostly with mRNA vaccine (RKI 2023d; 2024). This increases vaccination coverage needed by 5%, e.g., to 75% for the original estimate.

## Conclusions

Pandemic progression is described here simply by taking a 7-day moving average over the number of daily new cases, and other related epidemiological characteristics. This reveals critical regions of exponential growth and decay, and leads directly to the diagnostic basic and time-varying reproduction numbers.

The peak of the all-important first wave of incidence is much lower than following ones. Comparison with COVID-associated fatalities suggests four-fold underreporting in the initial wave, which reduces to three-fold for the second wave. Correcting for this difference, the wave-2 peak is $2.6\times$ that of wave-1, some of which might be attributed to a seasonal effect – because the first wave starts late. However, it remains unclear by how much initial preventative measures taken are effective in depressing the first peak.

Exponential rate constants of growth and decay, $r$, are more reliable indicators than is straight incidence, because they depend (via the logarithm) on ratios of 7-day average numbers of new cases (see, e.g., Fig. 1). Similarly also do reproduction numbers $R_0$ and $R_t$, either directly in Eqs. 7, 8, or indirectly via $r$ (Eqs. 11, 13-16). Rates and reproduction numbers afford a very direct approach to following progress of the pandemic, the effects of interventions, the evolution of new dominant viral variants, prediction of likely outcomes, and estimating necessary extents of vaccine coverage.

For numerical estimates of reproduction numbers, ideally we need the distribution of generation times. As already explained, we use instead serial intervals deduced directly from chains of infection for individuals displaying symptoms. COVID infections also include pre-symptomatic transmission, which complicates the fitting of probability distribution functions. The approach used here for Gaussian distributions, viz. Eq. 14, yields not such low values as those from the standard Eq. 15, and is consistent with direct application of the renewal equation (Marsh, 2025).

To get the basic reproduction number $R_0$, we use SI distributions from the beginning of the epidemic, because they are unmodified by later progression of the disease and introduction of preventative interventions. We retain the same distributions for subsequent instantaneous values of $R_t$, to allow direct comparison, whereas more realistic shorter values (cf., Ali et al., 2020; Park et al., 2021) would attenuate peaks and troughs in $R_t$ further relative $R_0$. Similar considerations apply also to $R_t$-values for new variants, which may have SIs different from the original SARS-Cov-2 and each other (Hart et al., 2022).

Mostly, we have assumed a homogeneous population, although this is not the case for people of different ages, and need not apply to different geographical locations. When normalized to individual populations, age differences in incidence are greatest for the first wave, and the reproduction number increases according to age (Fig. A.3). For subsequent stages, differences with age are mostly less pronounced, particularly in reproduction number. COVID-associated deaths increase exponentially with age from 20-25 years onwards (Fig. 10). Such age profiles can be used as a label in statistical analysis, to link infection prevalence and fatality, including across countries (O'Driscoll et al., 2021).

Initial rates $r_0$ are important in this approach because they lead to the basic reproduction number $R_0$. Unfortunately, they are subject to considerable variability (cf. Fig.

5). Case numbers are low at the start of the epidemic, as is the extent of testing, and reporting delays are likely to vary more. It therefore would help to combine case numbers with other indicators, such as hospitalization, intensive care occupancy, antibody seroprevalence, and virus count in waste water. Here we compare with COVID-associated deaths. Of course, statistical analysis of the links is invaluable, as are modeling studies (e.g., O'Driscoll et al., 2021; Khailaie et al., 2021).

**Acknowledgement**

I gratefully thank Christian Griesinger and the Department of NMR-based Structural Biology for essential support.

**Table 1.** Phases of the COVID-19 pandemic in Germany (Schilling et al., 2021; Tolksdorf et al., 2022).

| phase | start-date (year-week)[a] | day[b] (1Mar=1) |
|---|---|---|
| 1st. COVID-wave (I) | 2020-09 | -1 |
|     2020 summer trough (a) | 2020-20 | 76 |
|     2020 summer trough (b) | 2020-30 | 146 |
| 2nd. COVID-wave (II) | 2020-39 | 209 |
| 3rd. COVID-wave (III): VOC alpha | 2021-08 | 363 |
|     2021 summer trough | 2021-23 | 468 |
| 4th. COVID-wave (IV): VOC delta (summer) | 2021-30 | 517 |
|     VOC delta (aut/wint) | 2021-39 | 580 |
| 5th. COVID-wave (V): VOC omicron BA.1 | 2021-51 | 664 |
|     VOC omicron BA.2 | 2022-08 | 727 |
| 6th. COVID-wave (VI): VOC omicron BA.5 | 2022-21 | 818 |

[a]Week-number is corrected because 7-day incidence in Schilling et al. (2021) is centred on Friday of the previous week.

[b]Friday of week in start-date column.

**Table 2.** Exponential rate constants $r_t$ (day$^{-1}$) at different stages $t$ in the progress of infection. From log-linear regressions in Figs.1, 2, 4 and 8.[a]

| rate-constant | stage | from reporting [range, day] | from onset [range, day] |
|---|---|---|---|
| $r_0$ | wave-I | 0.264±0.007 [5–16] | 0.260±0.008 [4–11] |
| $r_1$ | | 0.072±0.002 [17–24] | -0.030±0.001 [17–33] |
| $r_2$ | | -0.058±0.001 [36–66] | -0.061±0.001 [35–59] |
| $r_3$ | | -0.040±0.001 [68–84] | -0.037±0.001 [60–84] |
| $r_4$ | 2020-smr. tr. b | 0.035±0.0003 [132–172] | 0.042±0.001 [127–143] |
| $r_5$ | wave-II | 0.074±0.001 [217–242] | 0.071±0.001 [216–233] |
| $r_6$ | | -0.030±0.0003 [314–343] | -0.025±0.0003 [324–347] |
| $r_7$ | wave-III, α | 0.036±0.0007 [371–388] | 0.040±0.0003 [369–379] |
| $r_8$ | | -0.029±0.001 [419–432] | -0.030±0.0005 [417–431] |
| $r_9$ | | -0.059±0.001 [435–468] | -0.060±0.0005 [432–461] |
| $r_{10}$ | | -0.083±0.005 [468–478] | -0.083±0.002 [462–474] |
| $r_{11}$ | 2021-smr. tr. | 0.064±0.002 [494–508] | 0.070±0.001 [487–499] |
| $r_{12}$ | wave-IV, δ | 0.061±0.001 [520–541] | 0.060±0.001 [518–531] |
| $r_{13}$ | aut./wint. | 0.046±0.002 [594–605] | 0.052±0.001 [589–596] |
| $r_{14}$ | | 0.037±0.001 [618–631] | 0.038±0.0005 [607–615] |
| $r_{15}$ | | -0.032±0.001 [646–655] | -0.037±0.001 [650–660] |
| $r_{16}$ | wave-V, BA.1 | 0.063±0.002 [671–684] | 0.041±0.001 [663–670] |
| $r_{17}$ | BA.2 | 0.030±0.001 [734–744] | 0.029±0.001 [726–735] |
| $r_{18}$ | | -0.043±0.002 [757–766] | -0.034±0.001 [755–763] |
| $r_{19}$ | wave-VI, BA.5 | 0.075±0.009 [825–834] | 0.056±0.003 [815–821] |

[a] The uncertainties (±) are standard errors in the slope SE = SD/√($\sum(t_i-\bar{t})^2$), where SD and $t_i$ are standard deviation of the fit, and day-numbers, respectively.

**Table 3.** Exponential rate constants $r_t$ for different variants of concern at the points of change-over between dominant variants. From log-linear regressions in Fig. 6 (top).

| variants | range (week) | crossing (day) | $r_{new}$ (day$^{-1}$) | $r_{old}$ (day$^{-1}$) | range (week) | $\Delta r$ (day$^{-1}$)[a] |
|---|---|---|---|---|---|---|
| alpha/original | 07-10 | 366 | 0.031±0.003 | -0.030±0.003 | 05-13 | 0.062±0.001 |
| delta/alpha | 23-25 | 476 | 0.084±0.014 | -0.059±0.007 | 23-26 | 0.135±0.007 |
| BA.1/delta | 51-01[b] | 671 | 0.079±0.026 | -0.066±0.003 | 50-02[b] | 0.146±0.008 |
| BA.2/BA.1 | 07-09 | 726 | 0.039±0.001 | -0.040±0.006 | 04-11 | 0.077±0.001 |
| BA.5/BA.2 | 22-24 | 830 | 0.050±0.010 | -0.060±0.001 | 21-24 | 0.125±0.010 |

[a] $\Delta r$ ($\equiv r_{new} - r_{old}$). From log-linear regressions in Fig. 6 (bottom).

[b] Year change: 2021 to 2022.

**Table 4.** Rate constants for new infections $r_I$, testing $r_N$ and positive tests $r$ (cf. Eq. 24). From log-linear regressions in Fig. 7.

| range (day) | <case>, $r$ (day$^{-1}$) | <test>, $r_N$ (day$^{-1}$) | <case>/<test>, $r_I$ (day$^{-1}$) | $r_N/r$ (%) |
|---|---|---|---|---|
| 33–61 ($r_2$)[a] | −0.039±0.002 | −0.008±0.002 | −0.031±0.003 | 21±6 |
| 215–243 ($r_5$) | 0.065±0.006 | 0.014±0.001 | 0.051±0.001 | 22±4 |
| 313–348 ($r_6$) | −0.023±0.001 | −0.004±0.001 | −0.019±0.001 | 17±5 |
| 369–390 ($r_7$) | 0.030±0.001 | 0.010±0.002 | 0.020±0.001 | 33±8 |
| 453–481 ($r_{10}$) | −0.064±0.004 | −0.011±0.001 | −0.053±0.003 | 17±3 |
| 495–530 ($r_{11-12}$) | 0.044±0.002 | −0.002±0.001 | 0.046±0.002 | −5±2 |
| 600–628 ($r_{13-14}$) | 0.047±0.001 | 0.025±0.003 | 0.022±0.002 | 53±8 |
| 670–691 ($r_{16}$) | 0.064±0.002 | 0.046±0.005 | 0.018±0.006 | 72±10 |
| 747–803 ($r_{18}$) | −0.028±0.001 | −0.021±0.002 | −0.007±0.001 | 75±10 |
| 817–845 ($r_{19}$) | 0.039±0.002 | 0.018±0.002 | 0.021±0.002 | 46±7 |
| 866–901 | −0.030±0.001 | −0.018±0.001 | −0.012±0.001 | 60±5 |
| 922–950 | 0.041±0.005 | 0.021±0.002 | 0.019±0.003 | 51±11 |

[a] Symbols in parentheses refer to corresponding exponential rate constants for reporting data in Table 2.

**Appendix 1**: Natural history of COVID-19 incidence data for Germany

The Robert-Koch Institute website (RKI, 2021a,f) provided four different time series for daily incidence of positive cases in Germany, namely those according to date of: i) official reporting to the local health authorities (*Gesundheitsamt*); ii) onset of symptoms, or otherwise of diagnosis, iii) receipt by RKI, and iv) onset of symptoms augmented by imputation of missing data and Nowcasting. Onset of symptoms relates most closely to the time of infection, but data do not cover all cases. Official reporting precedes date received by RKI; however the latter data is more immediate. The ECDC also provided daily incidence data (ECDC, 2020); from 26 Mar 2020 this source is identical to the RKI received-date daily listings (i.e., to iii). Johns Hopkins University CSSE dashboard (Dong et al., 2020) provided immediate data obtained by scanning rapidly published sources.

Figure A.1 shows the profiles of daily cases averaged over a 7-day window for the different time series, excepting ECDC. Profiles using onset-date (viz., ii and iv) clearly precede all others, and that for RKI official reporting (viz., i) precedes the remainder, particularly in the earlier stages. The JHU profile is more irregular, even after 7-day averaging. As expected, weekly modulation is largest for daily incidence according to official reporting date. The fitted amplitude of absolute-sine modulation is 88±3% of mean incidence, as compared to 66±4% for incidence according to RKI receiving date (68±4% for ECDC data). Fitted modulation amplitudes according to onset of symptoms are much smaller: 24±2% (23±2% for imputed). JHU data, on the other hand, does not conform with the regular modulation of an absolute-sine function.

Table A.1 lists rate constants $r_t$ for selected exponential regions in the time series of Fig. A.1.

Fig. A.1 also includes time series for daily occurrence of COVID-associated deaths. For RKI, fatality data correspond to received date (i.e., option iii above), where again ECDC data is identical from 26 Mar 2020 onwards. Data reported by RKI according to week of death (solid circles) are also included; this fatality time-series precedes those according to received (or reporting) date.

**Appendix 2**: Weekly modulation of COVID-19 incidence in Germany

The top panel of Fig. A.2 shows the amplitude of the weekly modulation that we get from the difference between daily cases and 7-day average, normalized to the latter. The data fall into three intervals, with stepwise increase in modulation amplitude between them. The initial region up to day-297 can be approximated by an absolute sine function (Eq. 21), shown by the solid line in Fig. A.2 with fitting parameters given in Table A.2. The second region extends from day-298 to day-750, at which point the modulation increases abruptly in amplitude with a midpoint around day-780 and transition width of ca. 20 days. This coincides with a large drop in the rate of testing at the end of the 2021-2022 winter wave of incidence, as seen in Fig. 7.

Table A.2 also includes parameters obtained from fitting the absolute sine to the modulation in the second region (day-298 to -750), and in the final region from day-837 onwards after the transition. The fit worsens in quality on proceeding to later intervals. For the first interval, the zero-level for weekly modulation, $f_o$, is close to the mean value of the absolute sine function, i.e., $2/\pi = 0.637$ times the maximum amplitude.

The lower panel of Fig. A.2 shows at which day of the week the minimum (open circles) and the maximum (solid circles) number of cases occurs. Almost without exception, the minima fall on Sundays. However, the maxima fall at different days within the week, and these groupings move from the end of the week towards the beginning, on proceeding from the first to third interval. This explains, at least partially, the deteriorating fits of an absolute sine function for the later intervals. With a minimum at the weekend, we expect an absolute sine to be maximum at the middle of the week. For interval 1, the maximum lies between Thursday and Wednesday; for interval 2, between Wednesday and Tuesday; for interval 3, first on Tuesday then ending between Tuesday and Monday. Bias towards the end of the week is expected from registration delays at the beginning of the epidemic. On the other hand, bias towards the beginning of the week could result from reductions in weekend manning at the end of the epidemic.

**Appendix 3:** Age-stratified reproduction numbers

Figure A.3 gives histograms according to age group for the basic ($R_0$, wave 1) and the instantaneous ($R_t$, waves 2-6) reproduction numbers, during 2020-2022. $R_0$ and $R_t$ are from Eq. 13, with SI-data from Bi et al., (2020). Incidence rates $r_t$ for the various epidemic stages are determined from Fig. 8.

**Table A.1.** Rate constants $r_t$, for exponential regions $t$ in differently reported time series of daily incidence in Germany. From log-linear regressions in Fig. A.1.

| dataset | $r_0$ (day$^{-1}$) | $r_2$ (day$^{-1}$) | $r_4$ (day$^{-1}$) | $r_5$ (day$^{-1}$) |
|---|---|---|---|---|
| RKIrpt | 0.263±0.008 [5,16] | -0.057±0.001 [36,67] | 0.035±0.0003 [132,172] | 0.073±0.001 [217,242] |
| RKIrcd | 0.239±0.009 [5,14] | -0.056±0.001 [38,68] | 0.034±0.0007 [153,174] | 0.073±0.001 [218,244] |
| RKIonset | 0.256±0.011 [5,11] | -0.061±0.001 [35,59] | 0.042±0.001 [127,143] | 0.063±0.001 [216,233] |
| RKINow | 0.273±0.009 [5,10] | -0.053±0.001 [35,60] | 0.041±0.001 [127,143] | 0.084±0.001 [223,234] |
| ECDC | 0.212±0.007 [4,16] | -[a] | -[a] | -[a] |
| JHU | 0.265±0.010 [7,17] | -0.058±0.001 [36,65] | 0.035±0.0005 [136,170] | 0.069±0.001 [217,247] |

[a]raw data identical to RKIrcd.

**Table A.2.** Parameters from non-linear least-squares fitting[a] Eq. 21 to modulation amplitudes over three adjacent non-overlapping time intervals.

| *range*: | [day-4,day-297] | [day-298,day-750] | [day-837,day-1185] |
|---|---|---|---|
| $A$ | 0.84±0.03 | 1.02±0.06 | 1.18±0.13 |
| $f_o$ | 0.632±0.008 | 0.639±0.007 | 0.618±0.049 |
| *wk* (day) | 7.002±0.002 | 6.987±0.001 | 6.998±0.003 |
| $t_o$ (day) | 0.95±0.04 | 1.80±0.03 | 1.05±0.09 |
| reduced $\chi^2$ | 0.013 | 0.025 | 0.168 |

[a]Levenberg-Marquardt algorithm implemented in OriginPro 2020 (OriginLab Corp.).

**Figure Legends**

Fig. 1. Progressive stages of the COVID-19 pandemic, correlated with daily incidence in Germany. Moving average of positive daily cases plotted against centre of the 7-day window. Day-1 is 1 Mar 2020. *y*-axis is logarithmic; straight lines mark regions of exponential growth.

Fig. 2. Daily number of new COVID-19 cases for Germany in 2020 (triangles); and 7-day moving average (circles). *Top*: according to reporting date; *bottom*: referred to date for onset of symptoms (*grey*: includes imputed missing cases). Day-1 is 1 Mar 2020; asterisks indicate public holidays; horizontal line: weekly incidence, 50 per 100,000 inhabitants. Data from Robert-Koch-Institute website (RKI, 2021a,b).

Fig. 3. Weekly modulation factor for daily new cases (circles). Difference between daily cases and 7-day average, normalized to the latter. Extended data from top panel in Fig. 2. *Solid line*: nonlinear least-squares fit of absolute sine function (Eq. 21).

Fig. 4. Daily number of new cases in 2020-2021, showing exponential regions: 7-day moving averages according to reporting date (circles) or onset of symptoms (diamonds; *grey*: includes imputed missing cases). Asterisks indicate public holidays; horizontal bars represent school summer holidays. Extended data from Fig. 2 with logarithmic *y*-axis (RKI, 2021a,b).

Fig. 5. Instantaneous $R_t$ (Eq. 7; open triangles) and cohort $R_t^C$ (Eq. 22; solid circles) reproduction numbers, 2020-2022. Deduced from 7-day averaged case numbers, based on reporting date (*top*) and onset of symptoms (*bottom*). Horizontal bars are values of $R_t$ for gamma- (Eq. 13, Bi et al., 2020: solid) and delta-function (Eq. 16 with $T_G = 4$ days: dotted) distributions of serial-interval; incidence rates $r_t$ are from Table 2. *y*-axes are logarithmic. Epidemic waves (w1-w6; Table 1) are indicated in the top panel, and changes in dominant viral variant (WT, $\alpha$, $\delta$, BA1, BA2, BA5) in the bottom panel.

Fig. 6. Variants of concern, 2021-2022. *Top panel*: relative populations (%) of original SARS-CoV-2 (open circles), Alpha-variant (B.1.1.7, solid squares), Delta-variant (B.1.617.2, open triangles), Omicron-variant (B.1.1.529 ≡ BA.1, solid inverted triangles), and Omicron-variants BA.2 (open diamonds) and BA.5 (solid left triangles). Weekly values for case data from random sampling of fully determined genomic sequences (RKI 2021c; 2023b), centred on Thursdays. *y*-axis is logarithmic; straight lines: linear regressions in common changeover ranges (see Table 3). *Bottom panel*: ratio of cases for newly dominating mutant relative to previous one, in changeover regions (Table 3). Alpha/original (solid squares), Delta/Alpha (open triangles), BA.1/Delta (solid inverted triangles), BA.2/BA.1 (open diamonds), BA.5/BA.2 (solid left triangles).

Fig. 7. Testing rate, 2020-2023. Number of tests reported ($N_{test}$, squares, left-hand scale), number of positive tests ($C_t$, circles, left-hand scale), and ratio of positive cases to total tests ($C_t/N_{test}$, triangles, right-hand scale). Daily 7-day averages from weekly totals are assigned to Thursdays of the week reported to RKI. Data from (RKI, 2021d; 2023c). *y*-axis is logarithmic; grey-shaded areas are common regions of exponential change.

Fig. 8. Age-dependent incidence, 2020-2022. *Top*: Daily cases for individuals aged 0-4 (right triangle), 5-14 (circles), 15-34 (up-triangle), 35-59 (down triangle), 60-79 (diamond), and 80+ (left triangle) years; all ages (squares). 7-day moving averages of new cases. *Bottom*: as for top panel, except that average daily cases, $\langle case \rangle_j$, are normalized to the fractional population, $N_j/N_{tot}$, in each age group, $j$. Data based on day of reporting (RKI, 2023a).

Fig. 9. COVID-associated deaths, 2020-2023. *Top*: Daily deaths predicted from daily infections *vs*. symptom-onset date $t_{os}$ (Eq. 25; solid circles). Predictions scaled to peak at day-40. Symptom-onset case data from Figs. 2, 4 (*grey*: includes imputed missing cases); probability density for onset-of-symptoms to death distribution from Verity et al. (2020). Open circles: COVID-associated deaths from weekly averages, centred on Thursdays (RKI, 2021b,e; and cf., 2025). *y*-axis is logarithmic. *Bottom*: case-fatality ratio, $cfr = \langle deaths \rangle / \langle predict \rangle$ from top panel.

Fig. 10. Age dependence of COVID-associated deaths. Age ranges: 0-4, 5-14, 15-34, 35-59, 60-79, over-80 years. Weekly deaths normalized to fractional population of age-group, averaged over reporting time periods shown (in days; day-1= 1 Mar 2020). *Solid circles*: total deaths in 2021, weekly averages normalized to fractional population of age-group. Data from RKI (2025) and SB (2025).

Fig. 11. Vaccination time course, 2021-2023. *Top*: Percentage of population twice vaccinated, or once vaccinated after recovery from infection, according to age groups: 60+ (diamond), 18-59 (inverted triangle), 12-17 (triangle), 5-11 (circle) years, or all ages (square). *Bottom*: Percentage of total population vaccinated once (square), twice (circle), three times (triangle), or four times (inverted triangle). Horizontal lines: percentage of total population in age groups (top to bottom): 18-59, 60+, 5-11, and 12-17 years. Dashed vertical lines correspond to those in the top panel. Data based on reporting to RKI (2023d).

Fig. 12. Vaccination target (% of total population) as vaccination proceeds. From Eq. 27, using cohort $R_t^c$ (Fig. 5) for daily cases based on reporting date (open circles) and on onset-of-symptoms (solid circles). *Squares*: % doubly vaccinated. Pandemic stages labelled with currently emerging dominant CoV-2 variant. *Horizontal line*: vaccination target for original variant.

Fig. A.1. 7-day moving averages for daily new cases (*upper*) and COVID-associated deaths (*lower*). *Open symbols*: according to RKI reporting date (o, RKIrpt), or onset of symptoms (□, RKIonset). *Crosses*: according to RKI received date (+, RKIrcd), or onset of symptoms augmented by imputation (×, RKIimpute). *Grey crosses*: Johns Hopkins CSSE data (×, JHU). Grey vertical lines are Sundays (day-1 = 1 Mar 2020); *y*-axis is logarithmic. Data from Robert-Koch Institute website (RKI, 2021a,e,f), and Johns Hopkins University CSSE website (Dong et al., 2020).

Fig. A.2. Weekly modulation factor for daily new cases, 2020-2023 (day-4 to -1185). *Top*: Difference between daily cases and 7-day average, normalized to the latter (circles and dashed line). Nonlinear least-squares fit of absolute sine function (Eq. 21), over range day-5 to day-259 (solid line). *Bottom*: day of the week for maximum (solid circles) and minimum (open circles) modulation.

Fig. A.3. Age-stratified reproduction numbers ($R_0$ or $R_t$) for epidemic waves 1-6 during 2020-2022 (see Table 1). $R_t$ is from Eq. 13, using SI-data from Bi et al., (2020). Corresponding incidence rates $r_t$ are determined from Fig. 8 over the ranges indicated above each group of bars. Shading indicates the age range for each bar.

**Figures**

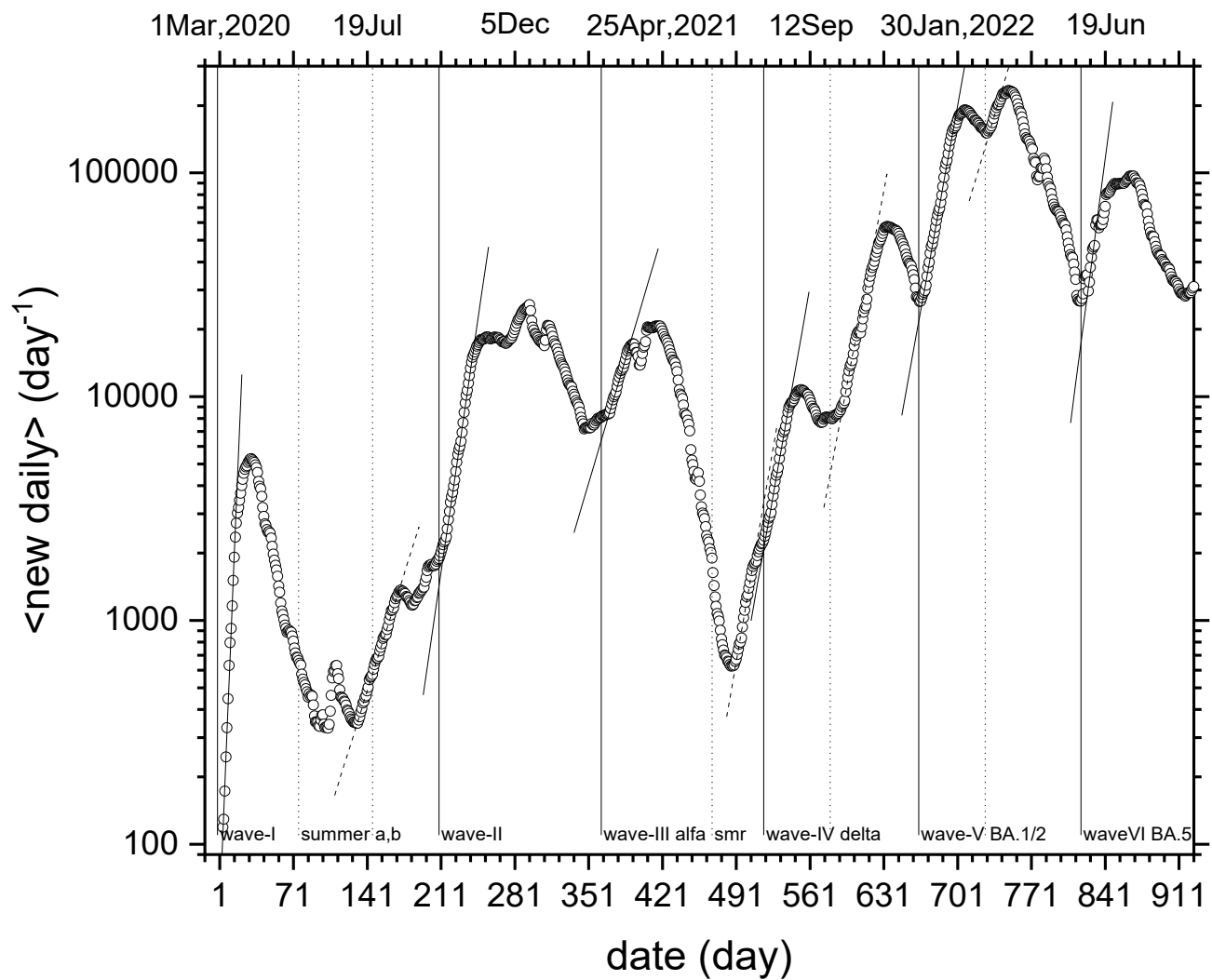

Fig. 1.

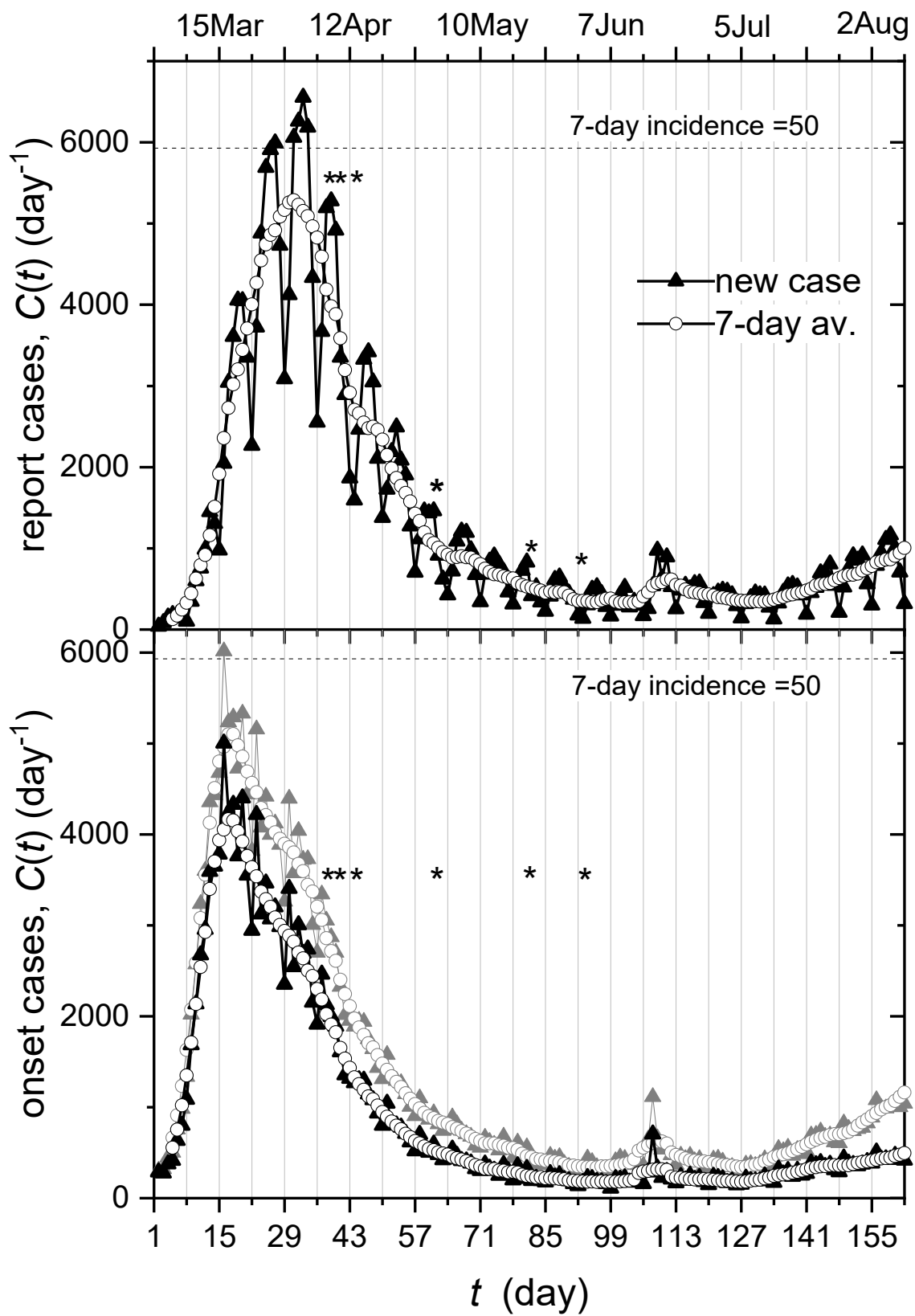

Fig. 2.

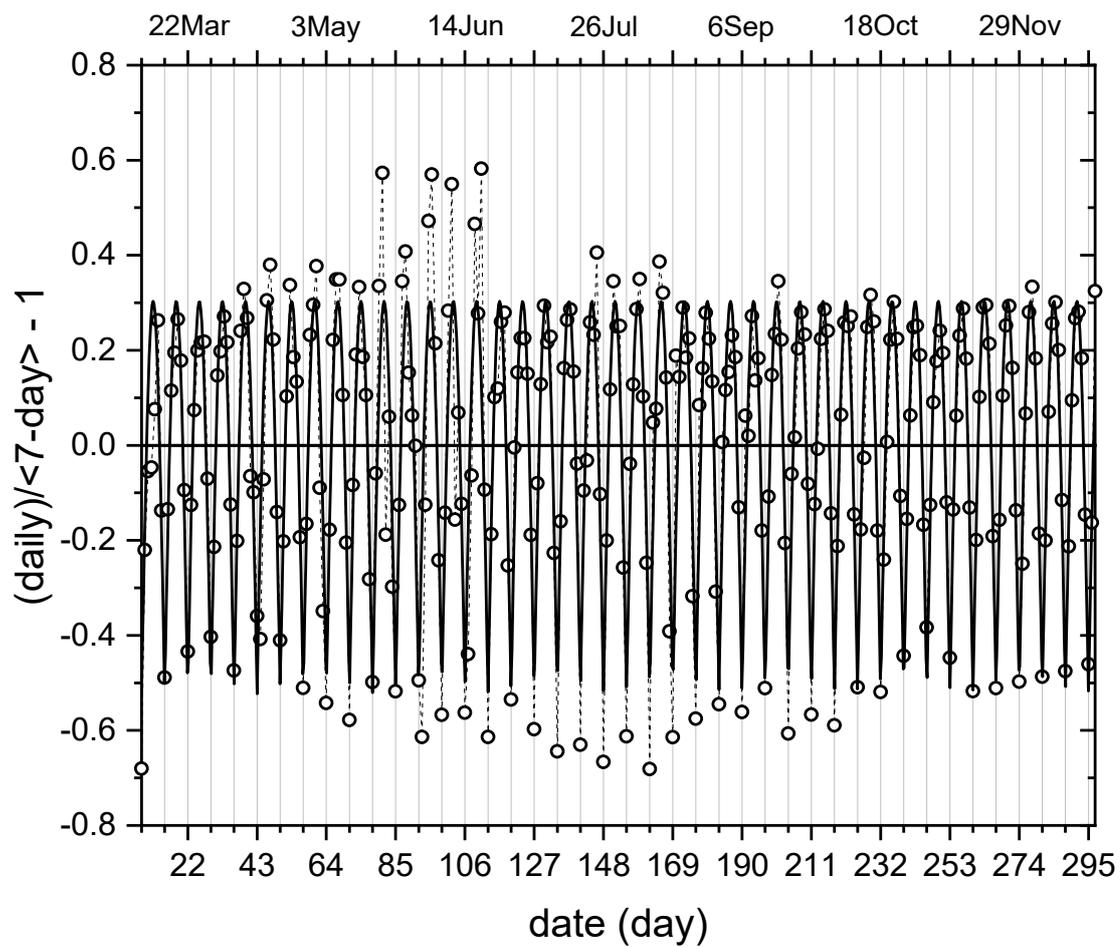

Fig. 3

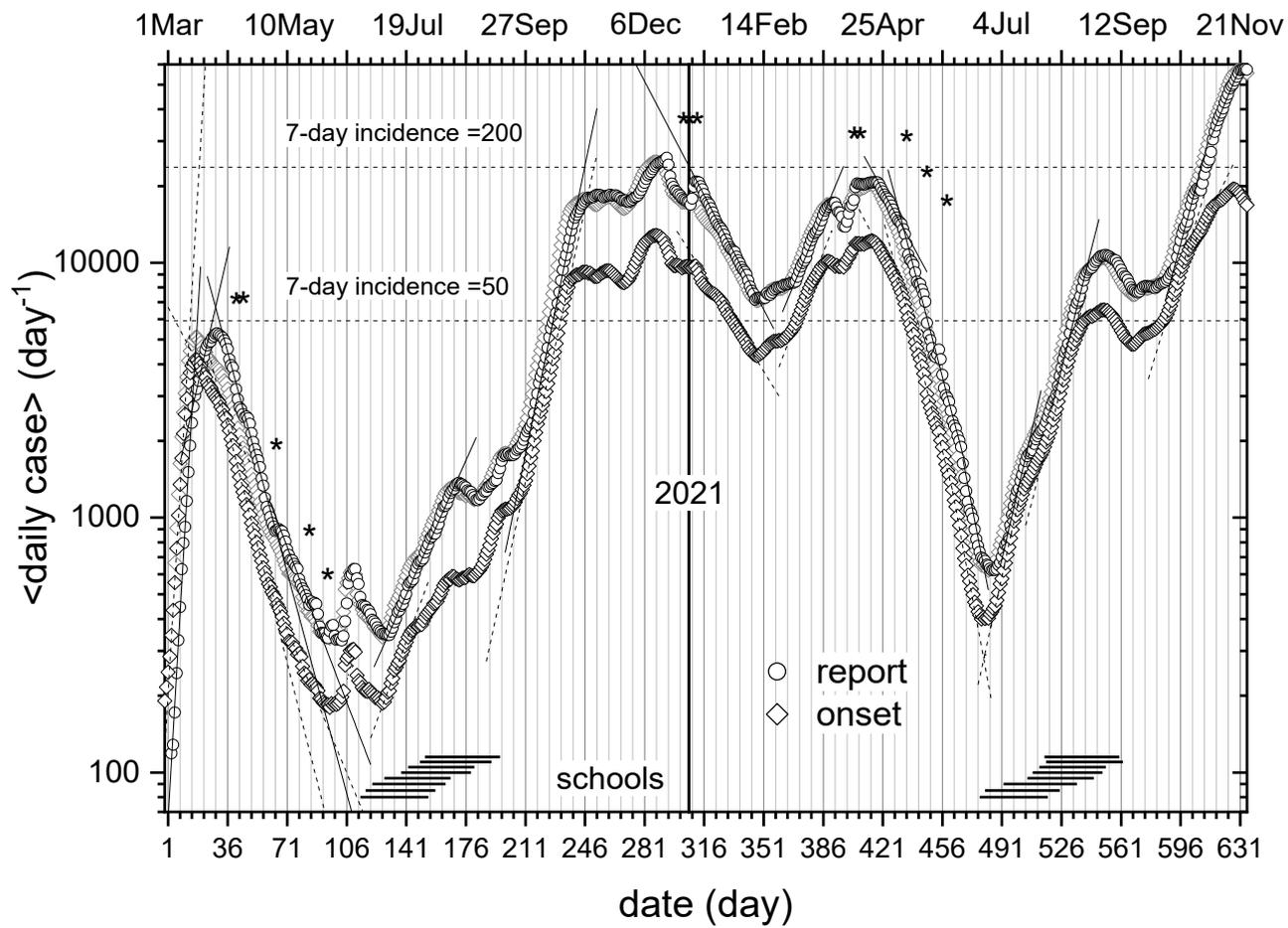

Fig. 4.

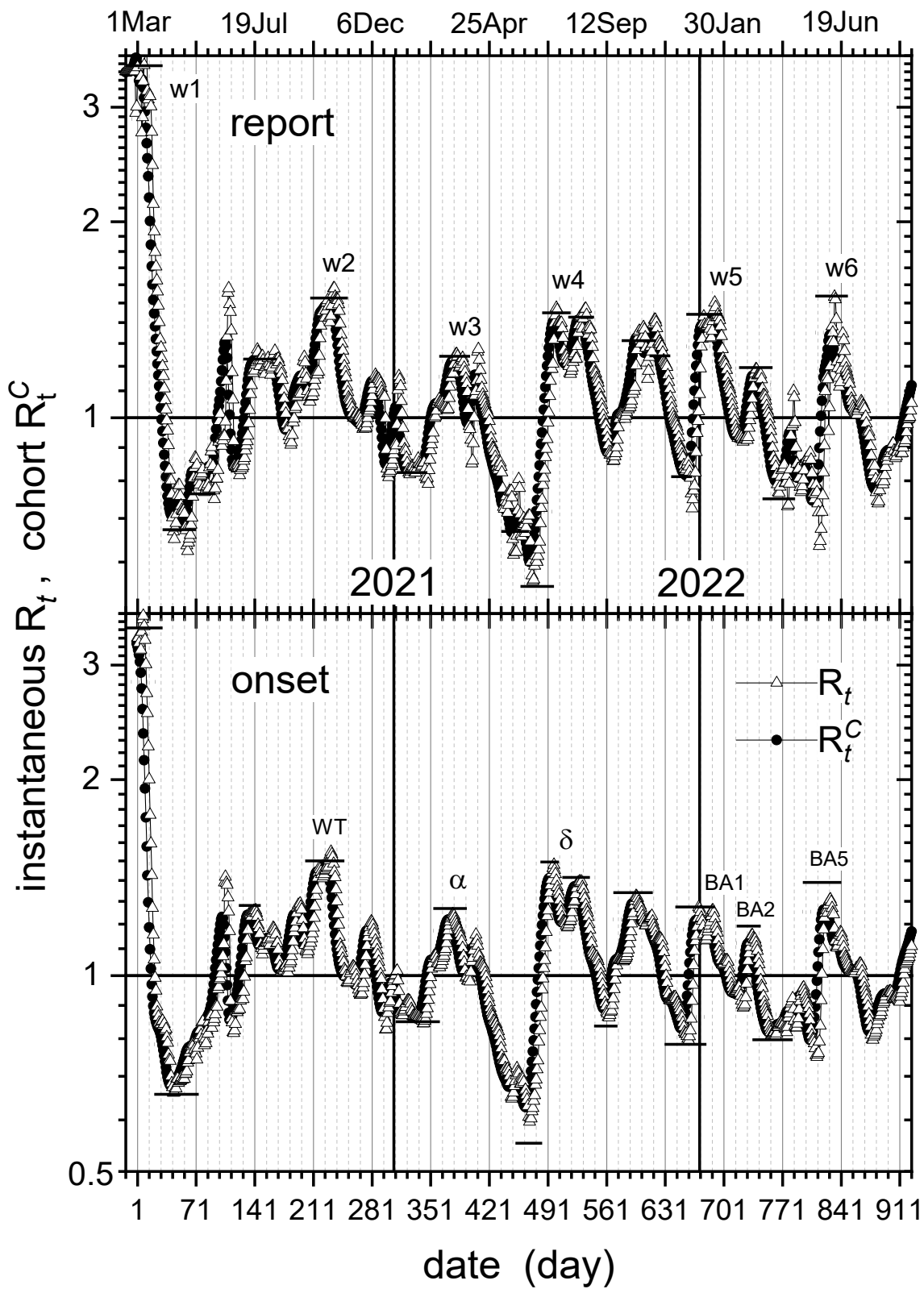

Fig. 5.

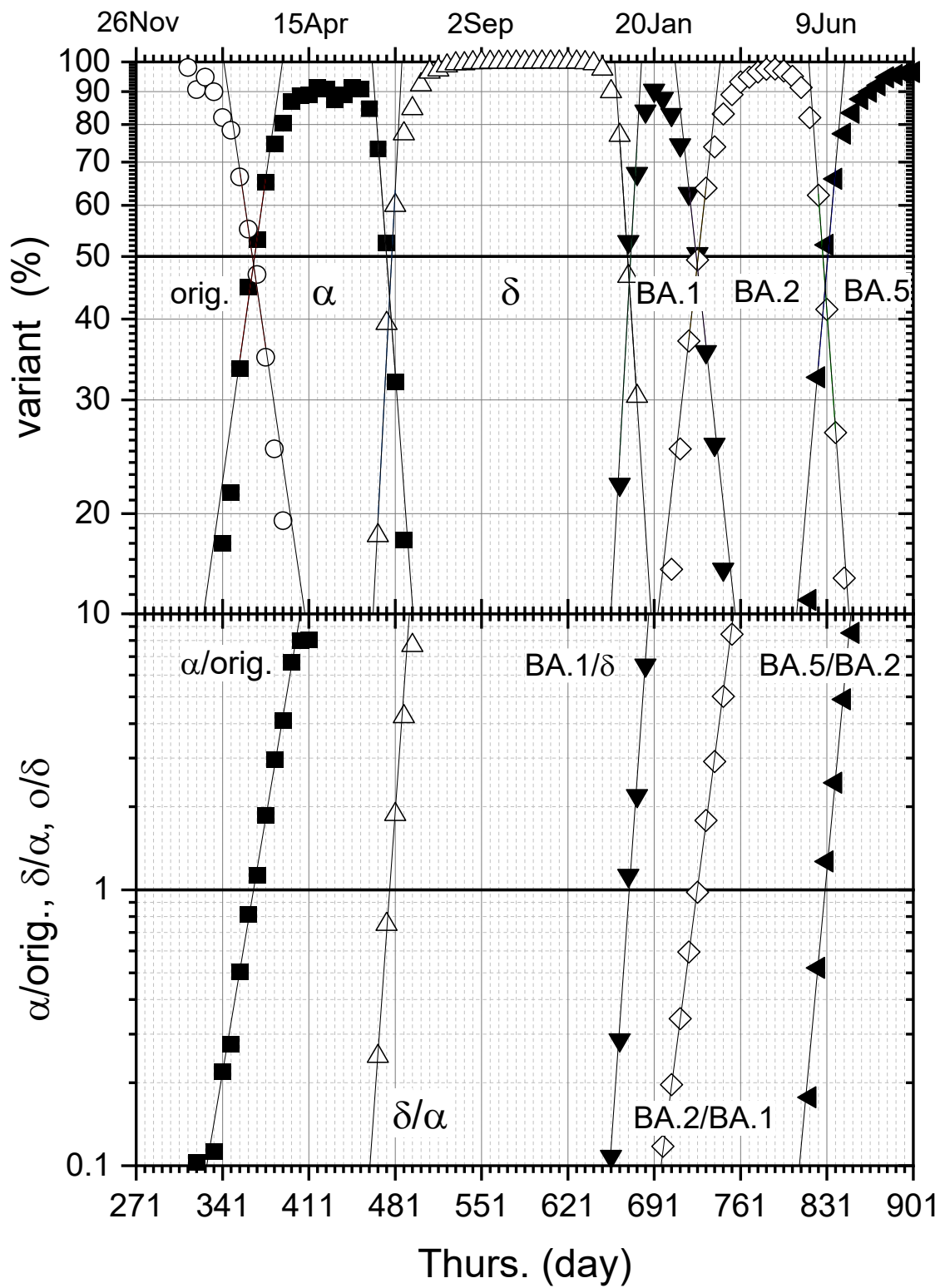

Fig. 6.

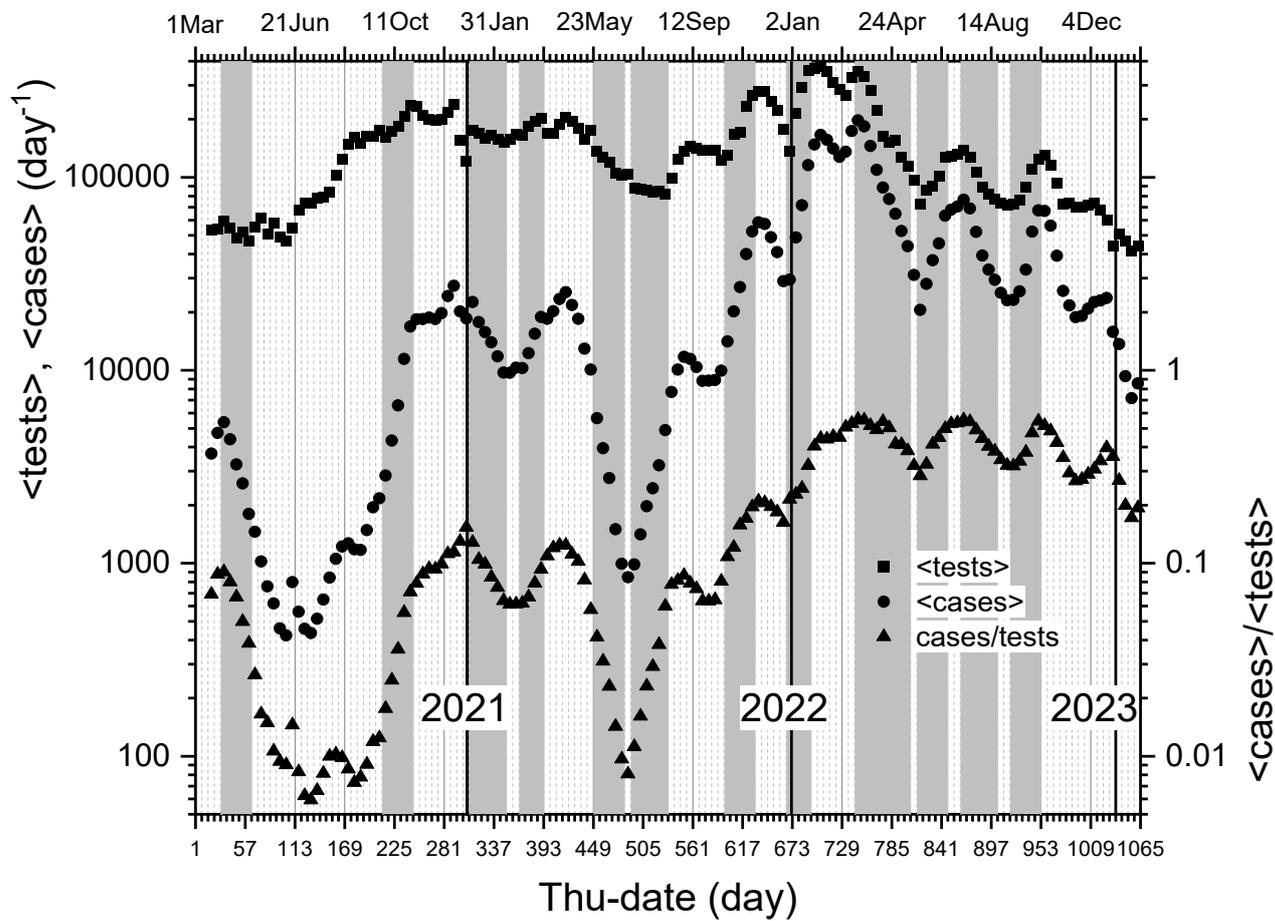

Fig. 7.

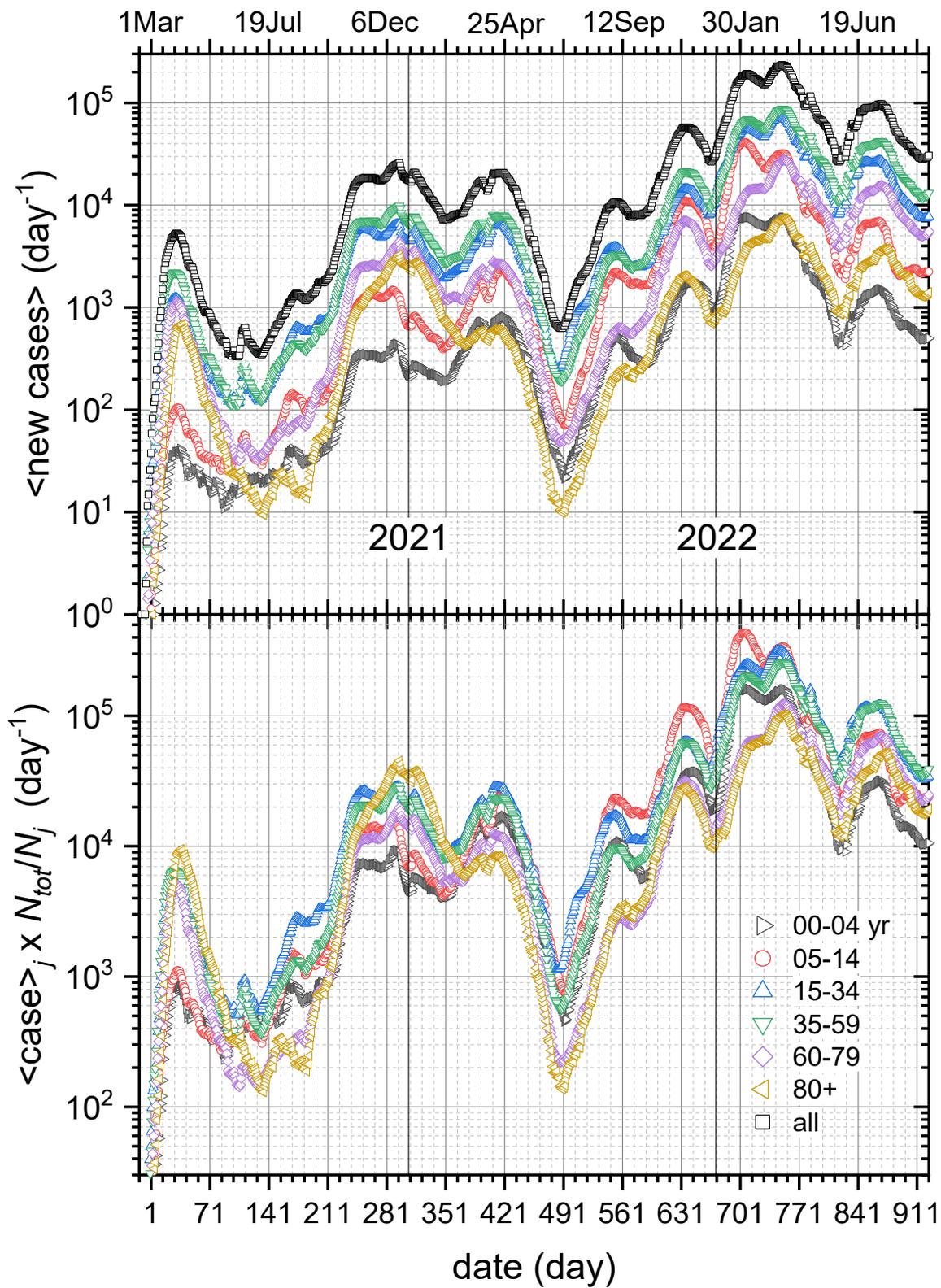

Fig. 8.

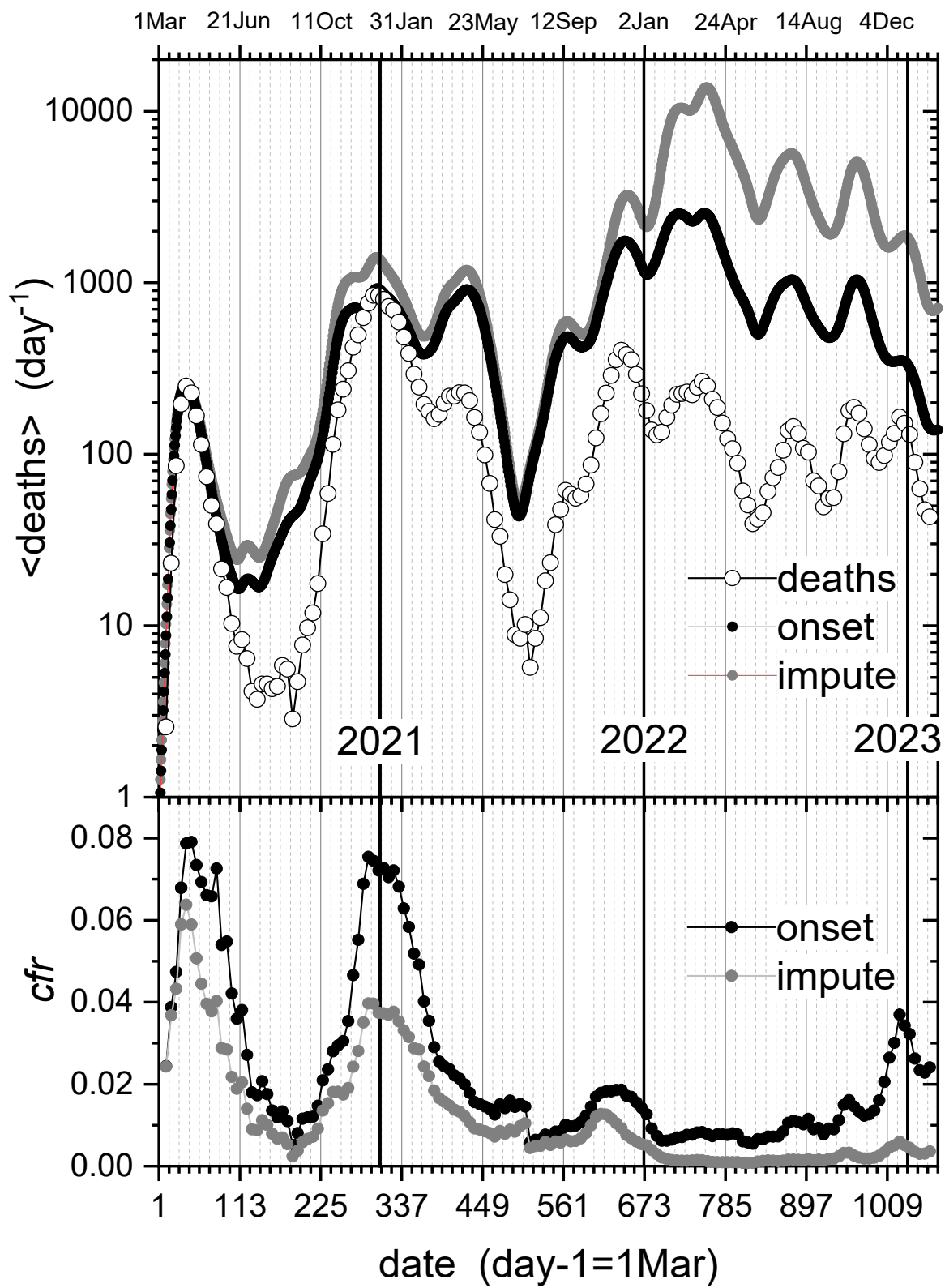

Fig. 9

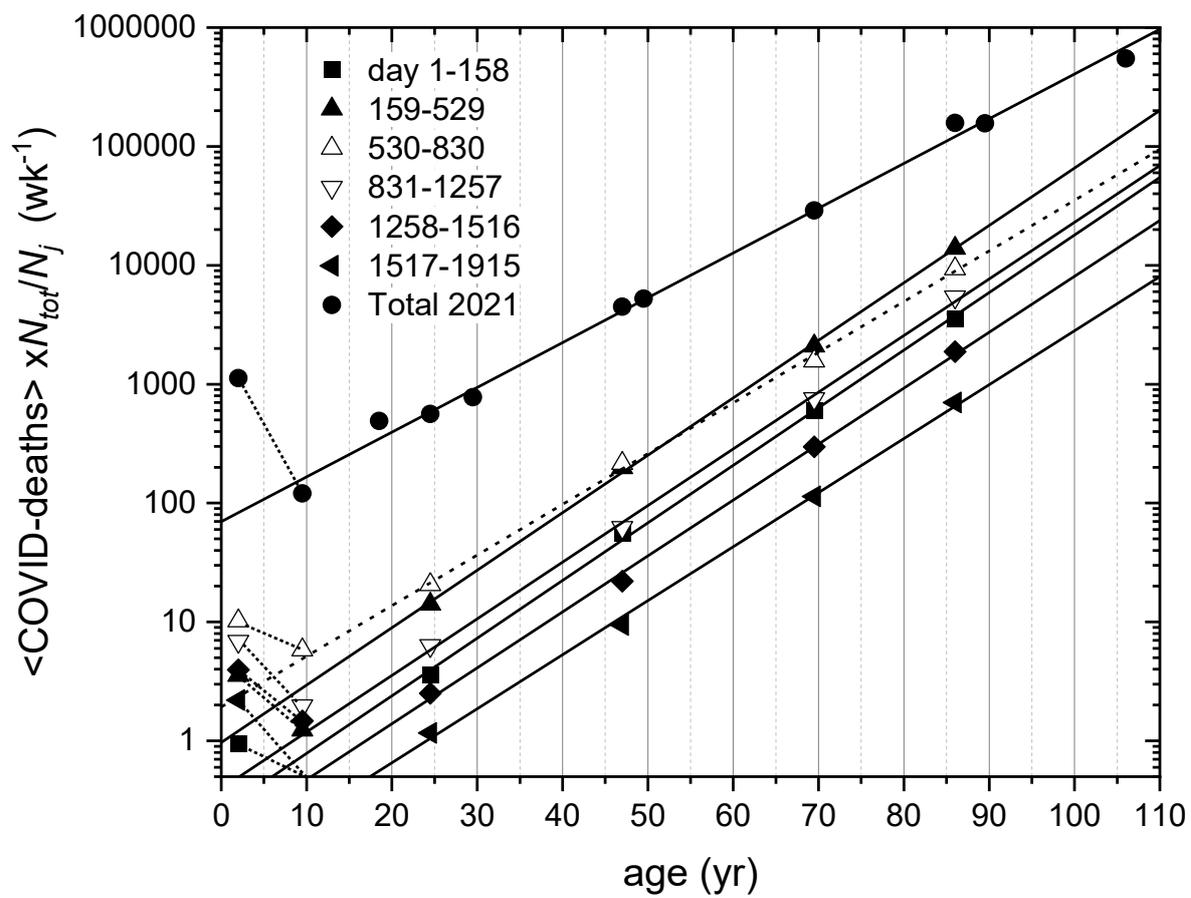

Fig. 10

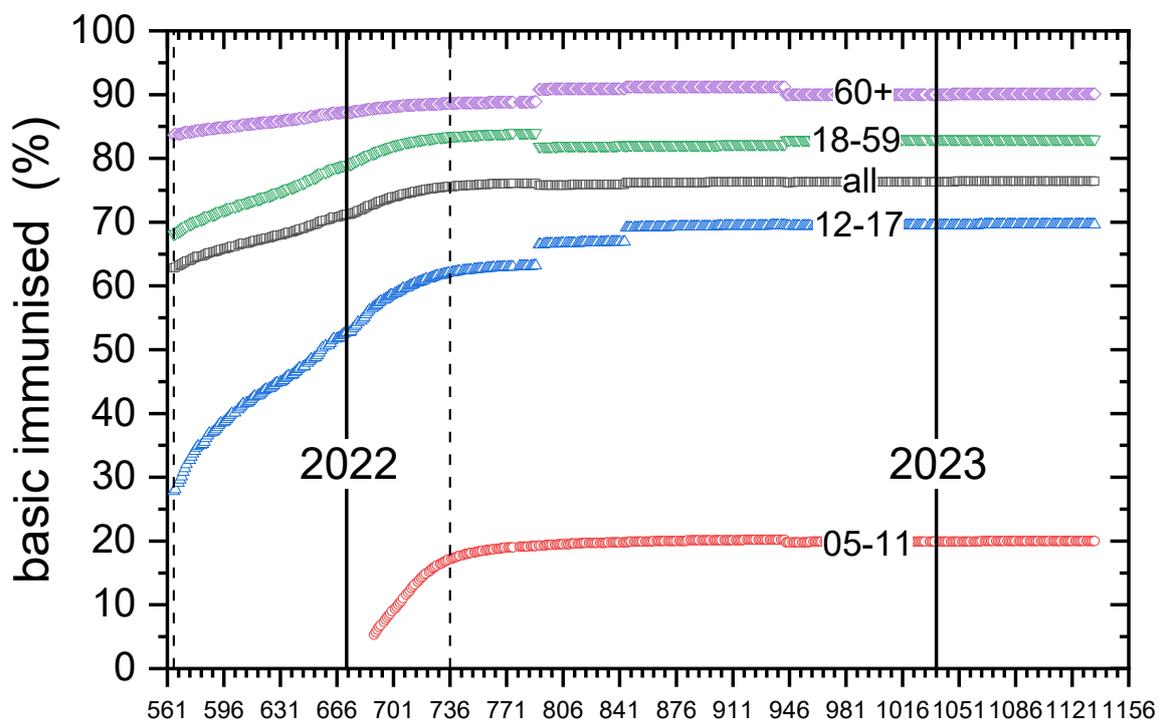

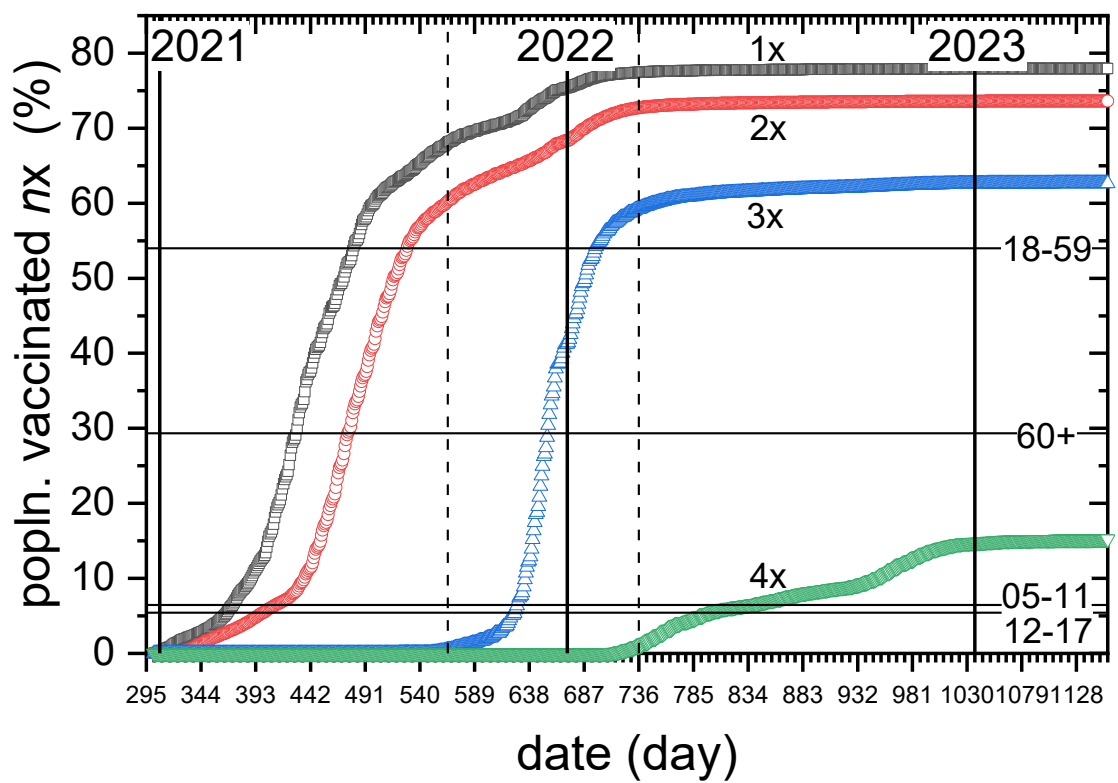

Fig. 11

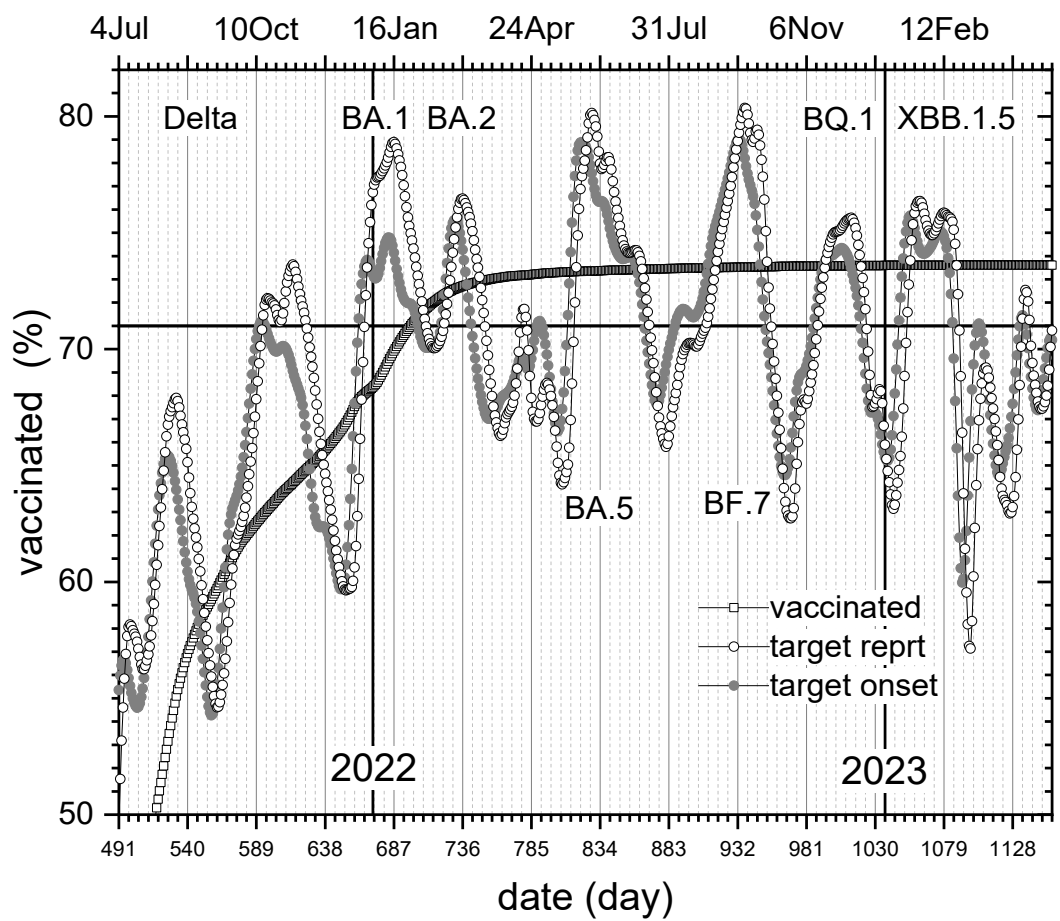

Fig. 12

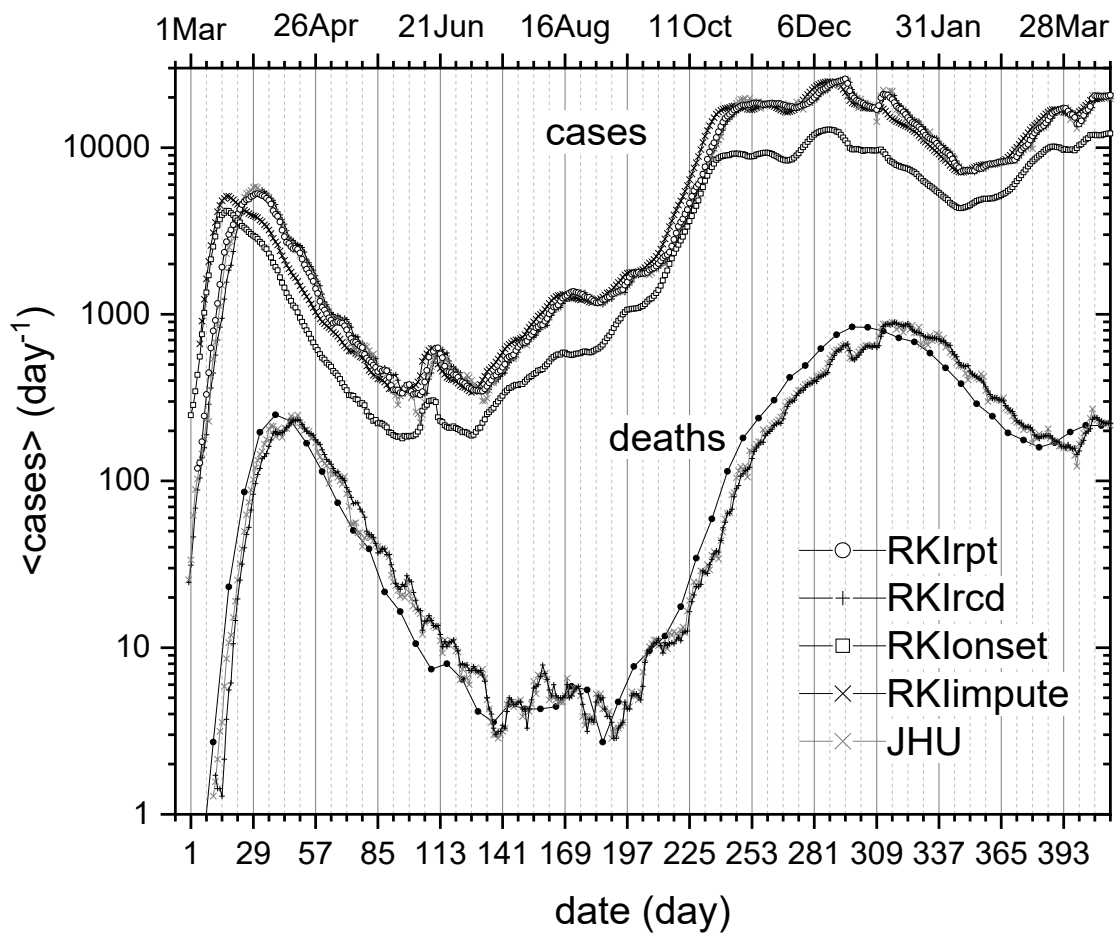

Fig. A.1

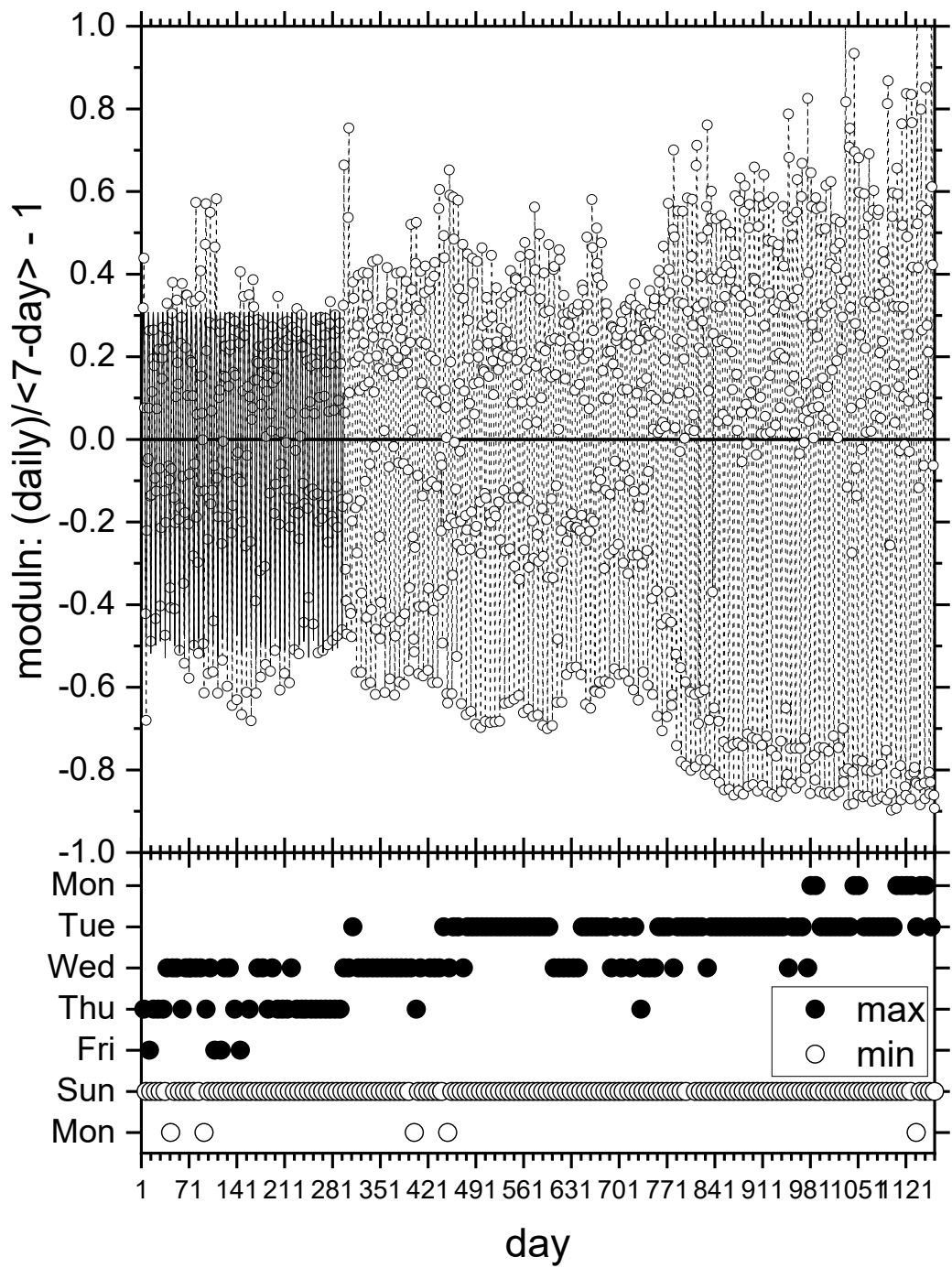

Fig. A.2

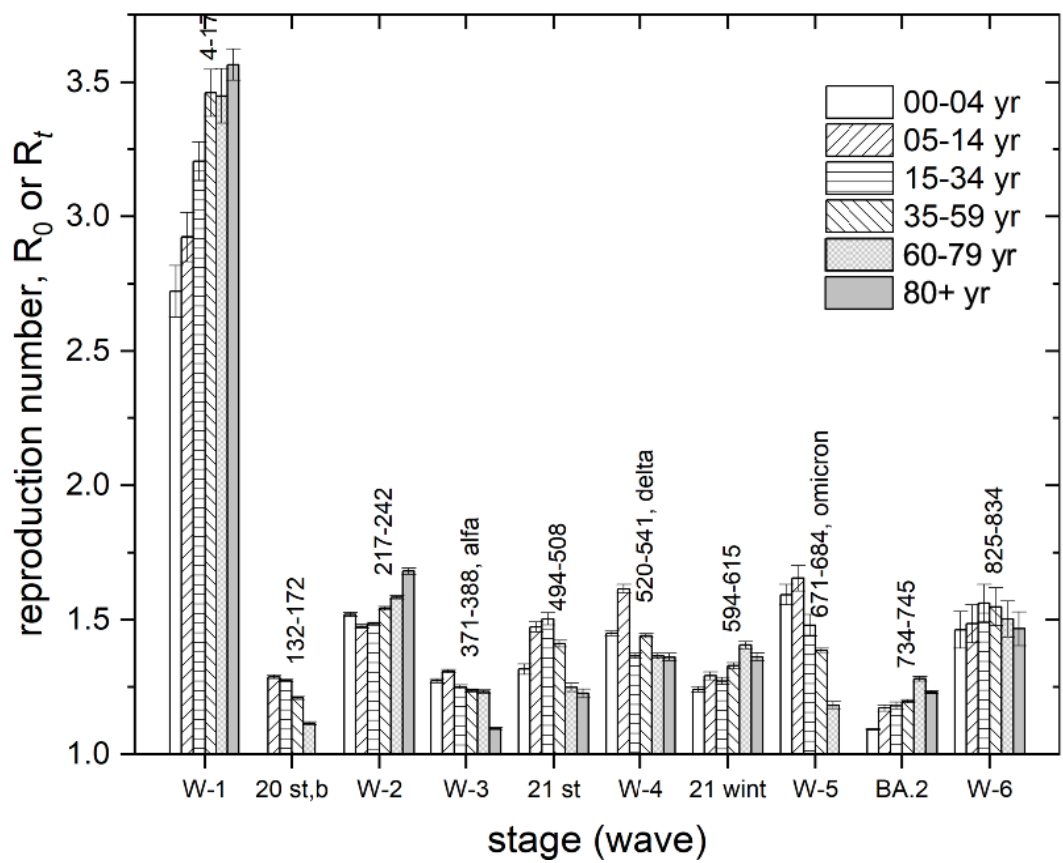

Fig. A.3.